\documentclass
[prb,twocolumn,showpacs,preprintnumbers,amsmath,amssymb,superscriptaddress,floatfix]{revtex4}
\usepackage{graphicx}
\usepackage{color}
\usepackage{dcolumn}
\usepackage{bm}

\def\IR{\relax{\rm I\kern-.18 em R}}
\def\btd{{\nabla}}

\def\boxeqnarray#1#2
{\fboxsep 1.ex\fbox{\parbox{#1}{\begin{eqnarray}#2\end{eqnarray}}}}

\def\IR{\relax{\rm I\kern-.18 em R}}
\def\btd{{\nabla}}

\def\region{\bigtriangleup_{\bf r}}
\def\Inc{\Delta}

\begin{document}
\preprint{generalCS -- A. Bad\'{\i}a-Maj\'{o}s, C. L\'opez and H. S. Ruiz}
\title{General critical states in type-II superconductors}
\author{A. Bad\'{\i}a\,--\,Maj\'os}
\email[Electronic address: ]{anabadia@unizar.es}
\affiliation{Departamento de F\'{\i}sica de la Materia
Condensada--I.C.M.A., Universidad de Zaragoza--C.S.I.C., Mar\'{\i}a
de Luna 1, E-50018 Zaragoza, Spain}

\author{C. L\'opez}
\affiliation{Departamento de Matem\'aticas, Universidad de Alcal\'a
de Henares, E-28871 Alcal\'a de Henares, Spain}

\author{H. S. Ruiz}
\affiliation{Departamento de F\'{\i}sica de la Materia
Condensada--I.C.M.A., Universidad de Zaragoza--C.S.I.C., Mar\'{\i}a
de Luna 1, E-50018 Zaragoza, Spain}

\date{\today}
%
%
\begin{abstract}

The magnetic flux dynamics of type-II superconductors within the critical state regime is posed in a generalized framework, by using a variational theory supported by well established physical principles. The equivalence between the variational statement and more
conventional treatments, based on the solution of the differential Maxwell
equations together with appropriate conductivity laws is shown. Advantages of the variational method are
emphasized, focusing on its numerical performance, that allows to explore new physical scenarios. In particular, we present the
extension of the so-called double critical state model to three dimensional
configurations in which only flux transport (T-states), cutting (C-states) or both mechanisms (CT-states) occur. The
theory is applied to several problems. First, we show the features
of the transition from T to CT states. Second, we give a
generalized expression for the flux cutting threshold in 3--D and show its relevance in the slab geometry. In addition, several models that allow to treat flux depinning and cutting mechanisms are compared. Finally, the longitudinal transport problem (current is applied parallel to the external magnetic field) is analyzed both under T and CT conditions. The complex interaction between shielding and transport is solved.

\end{abstract}
%
%
\maketitle

%
%
\section{Introduction}

The investigation of the macroscopic magnetic properties of type-II
superconductors in the mixed state is already a classical subject.
The essential physics behind the collected vast phenomenology has
been well known for decades,\cite{campbell_evetts,brandtRPP} and may
be basically analyzed in terms of interactions between the flux
lines themselves (lattice elasticity, line cutting), and
interactions with the underlying crystal structure (flux pinning).

For many purposes, it happens that the mesoscopic description may be
further simplified by means of appropriate material laws for the
coarse-grained fields, i.e.: magnetic induction ${\bf B}\equiv
\langle{\bf b}\rangle$, electric current density ${\bf J\equiv
\langle{\bf j}\rangle}$ and electric field ${\bf E\equiv \langle{\bf
e}\rangle}$. Averages are supposed to be taken over a volume containing a big enough number of vortices. Along this lines, a brilliant proposal, the so-called
critical state (CS) theory, originally introduced by C. P.
Bean\cite{bean} has been widespread used. Such a model allows to
capture the main features of the magnetic response of
superconductors with pinning at low frequencies and temperatures,
through the minimal mathematical complication. In its simplest form,
the CS theory involves to solve Amp\`ere's law $dB/dx=\mu_{0}J$ with
some prescription for the current density ($J=\pm J_{c}$ or $0$),
and under continuity boundary conditions that incorporate the
influence of the sources. Being a quasi-stationary approach, the CS
is customarily stated without an explicit role for the transient electric
field and the remaining Maxwell equations. Nevertheless, the
recognized prediction power of the  theory is not accidental.
Although veiled, the role of $E$ and Faraday's law is of great
importance for its soundness. Thus, we recall the
above statement may be related to an almost vertical $E(J)$ law,
idealized by the graph $E = 0$ for $J<J_c$ and $E\to\infty$ for
$J>J_c$. Here, $E$ stands for the induced electric field owing
to variations of the flux density, and Faraday's equation (in fact
Lenz's law) is implicitly used by selecting the actual value $\pm
J_{c}$ or $0$ that minimizes flux variations, when solving
$dB/dx=\mu_{0}J$ along the process. 

In the case of ideally
one-dimensional problems, i.e.: infinite cylinders and slabs in
parallel field configuration, the previous statements lead to the
prediction of the observable physics without ambiguity. As recalled,
one is making a correct use of the Maxwell equations for a conducting system
that experiments a sharp transition in terms of the current density.
Physically, the material law relates to the vortex pinning
phenomenon, that is macroscopically described by the average pinning
force constraint $|{\bf J}\times{\bf B}|\leq F_p$. Straightforwardly, this gives place to the depinning threshold limitation $J_{\perp}\leq J_{c\perp}(B)$. In
addition, one has to consider a high flux flow dissipation when the
limitation is exceeded (as related to the mathematical condition
$E\to\infty$). We call the readers attention that for the simplified geometries, the equality ${\bf J}={\bf J}_{\perp}$
(${\bf J}$ is  locally perpendicular to $\vec{B}$) is automatically
fulfilled.

The extension of the above ideas to non-idealized systems is not
a closed subject yet, it is of utter importance for the understanding of the experimental facts, and constitutes the main motivation of this
article. In brief, and following the spirit of Bean's model, we
consider the question of identifying a theory that allows a general
description of the low frequency electrodynamics of hard
superconductors with the least conceptual and mathematical load.

From the physical point of view, the allowance of non-parallel flux
lines leads to consider a new threshold for the current density,
now related to the disorientation of adjacent vortices. This was already
remarked by J. R. Clem and A. P\'erez-Gonz\'alez\cite{dcsm} who
analyzed the relation $J_{\parallel} = H d\alpha /dx$ with
$\alpha$ the angle characterizing the flux lines
orientation at a given depth within a superconducting slab. Thus,
stemming from the fact that a maximum angle gradient is allowed between
vortices so as to avoid cutting and recombination, one has to
consider $J_{\parallel}\leq J_{c\parallel}(B)$.\cite{brandtjltp79,clemyeh}
Remarkably, the above mentioned authors showed that both the flux
depinning and cutting effects may be treated in a generalized CS
framework. The upgraded theory ({\em double critical state model} or
DCSM) has been applied with high success since the early 80's for
the understanding of many experiments in which both $J_{\parallel}$
and $J_{\perp}$ are involved. In addition to the mentioned static
threshold conditions, the DCSM generalizes the one-dimensional
$E(J)$ graph for quasi-stationary processes in terms of the natural
concepts $E_{\parallel}(J_{\parallel})$ and $E_{\perp}(J_{\perp})$.

From the mathematical point of view, the DCSM has been mainly
applied to experiments with rotating magnetic field
components, still parallel to the surface of large samples, i.e.:
two-component fields are allowed, but only one independent variable
is considered. However, some aspects that have appeared linked to
the investigation of new materials (small crystallites and
high-T$_c$ films), as well as the refinement of previous studies
require more specialized statements, so as to include finite size
effects, sample inhomogeneity, anisotropy, etc. Along this line,
some recent advances have to be quoted. On the one side, it has been
shown that finite size effects for thin samples may be treated by
composition of quasi one-dimensional
statements.\cite{brandt_unusual} In principle, this idea would allow
to include both the $J_{c\perp}$ and $J_{c\parallel}$ limitations
but, to the moment, it has been exploited in the limit
$J_{c\perp}\ll J_{c\parallel}$. On the other hand, truly two-dimensional
configurations in which some symmetry property allows to assume
${\bf J}\perp {\bf B}$ have also been solved by numerical
methods.\cite{vanderbemden,badia07} More recently, as a remarkable
advance to be mentioned, the mathematical structure of the DCSM
solution in the three dimensional case has been described and
obtained variationally for some examples.\cite{kashima}

In this work, new perspectives on the application of variational
methods\cite{badiaPRL} will be presented. Contrary to some recent claim
about their restricted scope,\cite{brandt_unusual} they will be
shown to be equivalent to the more conventional differential
equation statements for solving CS problems. Moreover, in many instances, our solution will be used to extend previous results toward unexplored
physical scenarios. In particular, the influence of the parameter
ratio $\chi \equiv J_{c\parallel}/J_{c\perp}$ will be quantified in
three dimensional systems. Also, allowed by the capability of the
theory, we introduce a new critical angle gradient threshold, that
generalizes the two dimensional concept $d\alpha /dx =
J_{\parallel}/H \leq K_{c}$. On the other hand, the so-called longitudinal transport problem, i.e.: a situation in which a magnetic field is applied along the direction of the transport current will be studied in a 3--D configuration. As a central physical result
of our paper, it is shown that the variational method naturally
distinguishes between the inductive and potential parts of the
background electric field in the CS problems. It will
be shown that the incorporation of the physical idea of the
direction of the electric field is straightforward by combining
Gauss' law and the variational statement.

The paper is organized as follows. In Sec.~\ref{secgcs}, the
physical background of the general CS concept is introduced. The
underlying approximations (\ref{mqs}), the validity of the
associated material law (\ref{matlaw}), and the justification of a
variational statement (\ref{varprin}) are described in detail. In
Sec.~\ref{application} we give a number of explicit examples related
to the application of the general CS theory to three dimensional
systems. Specifically, we consider various magnetic processes for an
infinite slab with a penetrating magnetic field of the form $(H_x ,
H_y , H_z)$, and described by different models hosted in our theory.
It will be shown that our results fully coincide with alternative
formulations, when comparison is allowed. Finally,
Sec.~\ref{conclusions} is devoted to discuss the main results of this work.

%
%
\section{General critical states: theory}
\label{secgcs}

This section is devoted to introduce the theoretical background that
justifies the critical state concept as a valid
constitutive law for superconducting materials. First, recalling
that the CS must be considered an approximation within the
magnetoquasistationary (MQS) regime of the time-dependent Maxwell
equations, we will discuss on the related limits for the whole set
of electromagnetic physical quantities. Then, we will present a thorough discussion about the representation of the CS as a
$\bf{J}(\bf{E})$ law for a perfect conductor with restricted currents,
in the MQS regime. Finally, the
variational statement of the CS is stated. We will show that the
variational principle introduced in previous work \cite{badiaPRL,badiaPRBvector} is fully
equivalent to the more usual formulation in terms of differential
equation statements, in time-discretized form. Along this line, some
recent claims about the limited scope of variational
formulations\cite{brandt_unusual} have to be reconsidered.
%

%
\subsection{MQS approximation and its consequences}
\label{mqs}

Let us first concentrate on the physical implications related to the
MQS approximation within the critical state theory. Recall that, in general, the dynamical
behavior of the macroscopic electromagnetic fields is determined by the
Maxwell equations accompanied by material constitutive laws,  ${\bf H}({\bf B})$, ${\bf D}({\bf E})$, and ${\bf J}({\bf E})$. Thus, Faraday's and Amp\`ere's laws
represent a coupled system of time evolution field equations
\begin{equation}
\label{eq:maxwell1}
\partial _t{\bf B} = - \nabla \times {\bf E}\quad , \quad
\partial _t{\bf D} =  \nabla \times {\bf H} - {\bf J} \, {\rm .}
\end{equation}
Taking divergence in both sides of each, and recalling integrability (permutation of space and time derivatives) leads to the additional conditions
\begin{equation}
\label{eq:maxwell2}
\partial_{t}(\nabla\cdot{\bf B})=0 \quad ,\quad
 \partial_{t} (\nabla\cdot{\bf D}) + \nabla\cdot{\bf J} = 0 \,
{\rm .}
\end{equation}
Within this picture of the electromagnetic problem, the remainig Maxwell equations can be interpreted as ``(spatial) initial
conditions'' for Eq.(\ref{eq:maxwell2}), that define the existence of conserved electric charges, i.e.:
\begin{equation}
\label{eq:maxwell3}
 \nabla\cdot{\bf B}(t=0) = 0 \quad ,\quad \nabla\cdot {\bf D} (t=0) = \rho (t=0)\, {\rm .}
\end{equation}

Eqs. (\ref{eq:maxwell1}), upon substitution of $\bf H$,  $\bf D$ and $\bf J$
through the constitutive laws,
and with appropriate initial conditions, uniquely determine the evolution profiles ${\bf
B}({\bf r},t)$ and ${\bf E}({\bf r},t)$.

%
%
\begin{figure}
\begin{center}
{\includegraphics[width=0.3\textwidth]{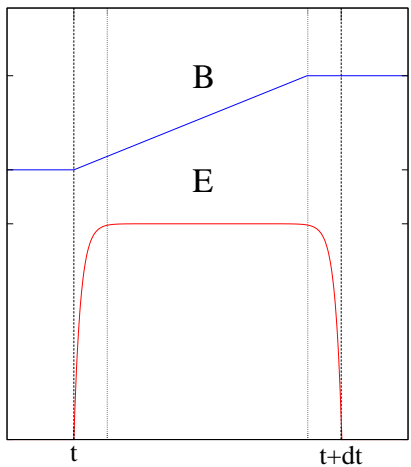}}
\caption{\label{fig_1}(Color online) Schematics of the time
dependence of the electromagnetic fields within the MQS
approximation. A ramp in the magnetic field is induced by the
external excitation, within the interval $[t,t+dt]$. As a
consequence of a very fast diffusion (elevated flux flow
resistivity), the electric field quickly adjusts to a constant value
along the interval. When the magnetic field ramp stops, $E$ goes
back to zero again. The {\em re-adjusting} vertical bands are
considered a second order effect and allow for charge separation and
recombination, according to the specific ${\bf E}({\bf J})$ model.}
\end{center}
\end{figure}

In this paper, as it is customary in hard superconductivity, we will consider
${\bf B} = \mu _0 {\bf H}$ as a valid approximation related to neglecting the equilibrium magnetization of the flux line lattice. On the other side, for {\em slow} and {\em uniform}
sweep rates of the magnetic sources, the
transient variables $\bf E$, $\bf D$ and $\rho$ are small, and proportional to
 $\dot{\bf B}$, whereas $\ddot {\bf B}$, $\dot {\bf E}$ and
$\dot {\rho}$ are negligible. Thus, the main hypothesis within the MQS regime is that
the {\em displacement} current densities $\partial _t{\bf D}$ are much smaller than $\bf J$ in the bulk, and vanish in a
first order treatment (see Fig. \ref{fig_1}). This causes a crucial change in the
mathematical structure of the Maxwell equations: Ampere's law is no
longer a time evolution equation, but becomes a purely spatial
condition. It reads
\begin{equation}
\btd \times {\bf B} \simeq \mu _0{\bf J} \,\, ,
\end{equation}
with approximate integrability
condition $\btd\cdot{\bf J}\simeq 0$.
In the MQS limit, Faraday's law is the unique time evolution equation.
Then, making use of the conductivity law
through its inverse function ${\bf E}({\bf J})$, one can find the
evolution profile ${\bf B}({\bf r},t)$ from
\begin{equation}
\label{eq:mqs}
\partial _t {\bf B} = - \btd \times {\bf E} \bigl({\mu _0}{\bf J} \simeq 
\btd \times {\bf B}\bigr) \, .
\end{equation}
We want to mention that the $B-$formulation in Eq. (\ref{eq:mqs}) is definitely the most extended one. However, the possibilities of $E-$formulations\cite{barret_prigozhin} and $J-$formulations\cite{wolsky} in which the dependent variables are the fields ${\bf E}$ and ${\bf J}$ have also been exploited by several authors. Also, a vector potential oriented theory ($A-$formulation) has been issued recently\cite{campbell}, that is a promising path for the investigation of 3--D problems.

Let's point out two relevant
consequences of the MQS limit.
\begin{enumerate}
\item  The constitutive law ${\bf D}({\bf E})$,
which is not used in Eq. (\ref{eq:mqs}), plays no role in the evolution
of the magnetic variables $\bf B$ and $\bf J$; the magnetic
``sector'' is uncoupled from the charge density profile because the
coupling term (charge recombination) has disappeared.
\item Only the inductive component
of $\bf E$ (given by $\btd \times {\bf E}_{\rm ind} = - \dot {\bf B}$, $\btd\cdot
{\bf E}_{\rm ind} = 0$) determines the evolution of $\bf B$
(Faraday's law). The conducting law in its inverse formulation ${\bf
E}({\bf J})$ presents some ambiguity, as far as two different material laws
related by ${\bf E}_2 ({\bf J}) = {\bf E}_1 ({\bf J}) +
\btd \Phi ({\bf J})$ determine the same magnetic and current density profiles.
\end{enumerate}

Going into some more detail, whereas for the complete Maxwell equations statement, the 
potential component of the electric field ($\btd \times{\bf E}_{\rm pot} = 0$, $\epsilon _0 \btd \cdot {\bf E}_{\rm pot} = \rho$), is coupled to $\bf B$ and  ${\bf E}_{\rm ind}$ through the $\dot {\bf D}$
term, within the MQS limit it is irrelevant for the magnetic quantities. 
In fact, one can include the presence of charge densities without contradiction with the condition ${\nabla}\cdot{\bf J}\simeq 0$ by means of inhomogeneity or non-linearity in the ${\bf E}({\bf J})$ relation. Then one has ${\nabla}\cdot{\bf J}= 0 \,\not\hspace{-1.ex}{\Rightarrow}\,{\nabla}\cdot{\bf E}= 0$. The charge density $\rho$ can be understood as a
parameterized charge of {\it static} character, as far as $\dot {\rho}$ is neglected. As indicated above, once integrated the magnetic variables, and computed $\btd
\cdot{\bf E}$, one has  the freedom to modify the electrostatic sector
if necessary, by the rule ${\bf E}({\bf J}) +
\btd \Phi$, while still maintaining the values of $\bf B$ and $\bf J$. This invariance can
be of practical interest, as far as the ``electrostatic'' behavior
in the CS is still under discussion, because of the inherent difficulties in
the direct measurement of transient charge
densities. Recent advances have to be quoted \cite{joos}, but they are still based on the analysis of ${\bf E}_{\rm ind}$ and some ansatz on the direction of the electric field. To be specific, ${\bf E}\parallel{\bf J}$ is assumed in that work.
%
%
\subsection{Material law: the critical state}
\label{matlaw}
Now, we will be more explicit about the material law ${\bf J}({\bf E})$
that dictates the magnetic response of a superconducting
sample in the critical state and for a given external excitation. For simplicity, we start with an overview of the  material law for 1--D systems (infinite slabs or cylinders with the external field applied along symmetry axis). The physical concepts will be eventually generalized to 3--D. 

\subsubsection{1--D Critical States}

%
%
%
\begin{figure}
\begin{center}
{\includegraphics[width=0.3\textwidth]{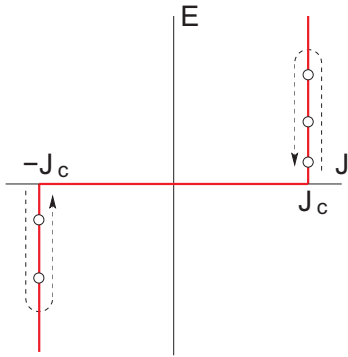}}
\caption{\label{fig_2}(Color online) Schematic representation of the
CS ${E}({J})$ model. The electric field arises when some {\em
critical} condition for the volume current density is reached ($J_c$
in this 1--D representation). Corresponding to the MQS approximation
in Fig. \ref{fig_1}, the electric field instantaneously increases to
a certain value, determined by the rate of variation of the magnetic
field, and then goes back to zero.}
\end{center}
\end{figure}
%

For our purposes, it is sufficient to recall that the basic structure of the CS relates to an
experimental graph within the $\{V , I \}$ plane that basically contains two regions:
\begin{enumerate}
 \item $-I_c \leq I \leq I_c$ with perfect conducting behavior, i.e.:
 $V=0$ and $\partial _t I = 0$.
\item For $I \gtrsim I_c$, the curve is characterized by a high  
${\partial_{I} V}$ slope (and antisymmetric for $I \lesssim - I_c$). Further steps,
with $I$ increasing above the critical value $I_c$, i.e., the eventual transition to the normal state, may be neglected for slow sweep rates of the external sources, which produce moderate electric fields.
\end{enumerate}
Within the local description level, different models have been used for the corresponding $E \leftrightarrow J$ graph, the most popular being
\begin{enumerate}
\item The {\em power law} model: $E=\alpha\;{\rm sgn}(J)\left({|J|}/{J_c}\right) ^n$, with $\alpha$ a constant and $n$ high.
\item The {\em piecewise continuous linear} approximation: $E=0$ for $|J|\leq J_c$, and $E=\beta\;{\rm sgn}(J)(|J|-J_c)$ for $|J|> J_c$, $\beta$ having a high value.

This model and the previous one present a small dependence on the sweep rate, as far as different values of $E$ give way to a slightly different $J$.
\item {\em Bean's model}: constant $J$ for $E=0$, and $J={\rm sgn}(E)J_c$ for $E\neq 0$ (see Fig. \ref{fig_2}).

This is the simplest model, without sweep rate dependence because only the sign of $E$ enters the theory.

\end{enumerate}
Bean's model captures the main features of the CS and has been widely used with very good performance since its proposal in the early sixties.\cite{bean} On the other hand, we notice that it may be obtained from the other representations: it corresponds to the limiting cases $n\to\infty$ and $\beta\to\infty$ respectively, as the reader can easily check. Considering the above ideas as the essential hypotheses of the CS theory, the 3--D critical state model will be formulated upon its generalization. The well known experimental evidence of a practical sweep rate independence for magnetic moment measurements (unless for high frequency AC sources or at elevated temperatures)
reinforces this simple model as a valid tool in the CS.

In some treatments, the first or second models are 
implemented, in order to transfer a full ${\bf E}({\bf J})$ law to the Maxwell equations.
On the other hand, being rate independent, Bean's model is no longer a time--dependent problem, but a path--dependent one, i.e.: the trajectory of the external sources
${\bf H}_{0}$ uniquely determines the magnetic evolution of the sample.\cite{path} This makes an important difference when one compares to more standard treatments, as far as Faraday's law is not completely determined from the path. Strictly speaking, one has

\begin{equation}
\label{eq:faradCS}
\Inc {\bf B} = - \btd \times [{\bf E} \Inc t] \,\, ,
\end{equation}

with $\Inc t$ (and therefore $|{\bf E}|$) gauged by the evolution of the external sources. In other words, the absence of an intrinsic time constant gives way to the arbitrariness in the time scale of the problem. On the other hand, the magnitude $|{\bf E}|$ is not relevant for the $\bf B$ and $\bf J$ profiles. In fact, in the applications of Beans's model, Faraday's law is not strictly 
solved. It is just the sign rule (the {\it vectorial}
part of the material law), that is used to integrate Ampere's law. Notice that such sign
rule corresponds to a maximal shielding response against magnetic vector variations, and thus,  determines the selection of $J= \pm J_c$. 

Notice that, by symmetry, in the 1--D problems one has
${\bf J}\parallel {\bf E}$ and both quantities are orthogonal to $\bf B$. Thus, at a basic level, the 1--D CS concept is grounded on the existence of pinning forces that act as a barrier against flux flow. The physical threshold related to a maximum value of the force balancing the magnetostatic term ${\bf J}\times{\bf B}$ gives place to the concept of maximum (critical) current density, and thus to the law
\begin{equation}
\label{eq:perp1D}
J_{\perp}={\rm sgn}(E_{\perp})J_{c\perp} \qquad{\rm for}\qquad E_{\perp}\neq 0\,\, .
\end{equation}
Here, $E_{\perp}$ stands for the component of ${\bf E}$ along the direction ${\bf B}\times ({\bf J}\times{\bf B})$.

\subsubsection{General (3--D) Critical States}

\begin{figure}
\begin{center}
{\includegraphics[width=0.45\textwidth]{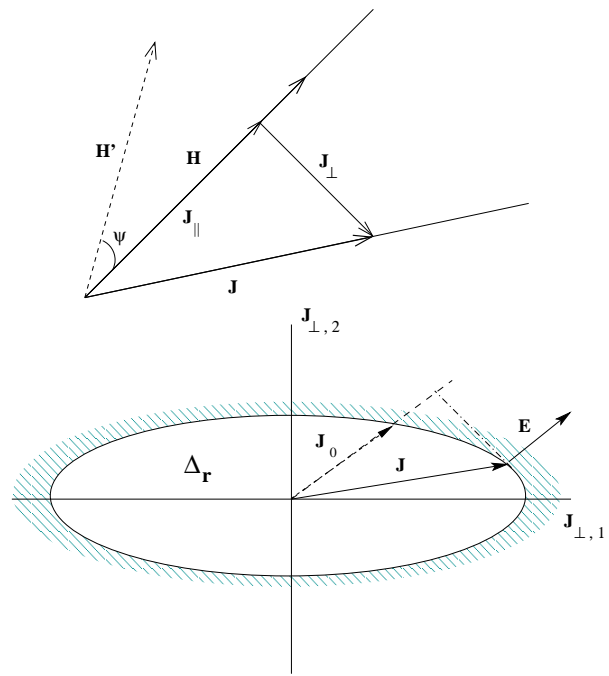}}
\caption{\label{fig_3} Top: schematic representation of the relative
orientations of the local magnetic field {\bf H} and electric
current density {\bf J}. The current is decomposed into its parallel
and perpendicular components, i.e.: ${\bf J}={\bf
J}_{\parallel}+{\bf J}_{\perp}$. Also sketched is the direction of
the magnetic field at some neighboring point, at an angle $\psi$.
The vectors ${\bf H}$, ${\bf H}'$ and ${\bf J}$ do not
necessarily lie at the same plane.
Bottom: the {\em perfect conducting} region within the plane perpendicular to the local magnetic field. An induced electric field is shown. Initially (${\bf J}_{0}$), the high dissipation region is touched, but almost instantaneously {\bf J} shifts along the boundary, reaching a point where the condition ${\bf E}\perp \partial \region$ is fulfilled. Anisotropy within the plane is allowed.}
\end{center}
\end{figure}

\begin{figure}[t]
\begin{center}
{\includegraphics[width=0.4\textwidth]{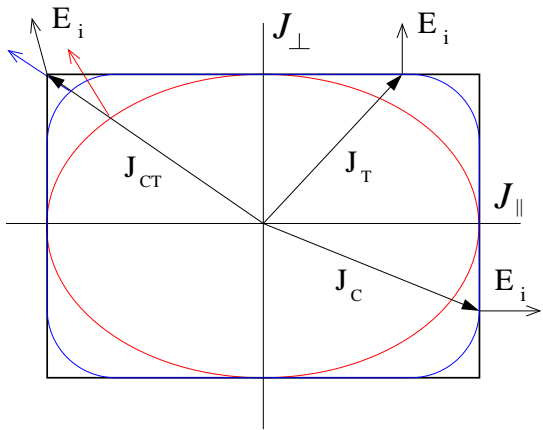}}
\caption{\label{fig_4} Geometric interpretation of the Critical
State behavior for the DCSM case. {\bf J} is constrained to the
boundary of a rectangular region. T, C and CT states are related to
the horizontal and vertical sides, and to the corners. Two models in which the corners of the DCSM region have been smoothed are also shown.}
\end{center}
\end{figure}

Let us now see how the above ideas may be translated to a 3--D scenario from the fundamental point of view. The main issue is that, in general, the parallelism of ${\bf E}$ and ${\bf J}$ and their perpendicularity to ${\bf B}$ are no longer warranted. Then, a {\em sign rule} does not suffice for determining the solution. A {\em vectorial rule} is needed and attention must be paid to its mathematical consistency, as well as to the physical significance. In previous work,\cite{badiaPRL,badiaPRBvector,badiaJLTP} we introduced a geometrical concept that may be of much help when discussing the idea of a general critical state theory. There must be a region $\region$ within the ${\bf J}$-space (possibly oriented according to the local magnetic field $\hat {\bf B}$,
and/or also depending on $|{\bf B}|$ and $\bf r$) such that non-dissipative current flow occurs when the condition ${\bf J}\in\region$ is verified (see Figs. \ref{fig_3}, \ref{fig_4}). This concept, together with a very high dissipation when ${\bf J}$ is driven outside $\region$ by some non vanishing electric field, suffice to determine the relation between the directions of ${\bf J}$ and ${\bf E}$. This may be done according to the following argument.

\begin{itemize}
\item[$\star$] In the {\em critical state}, the forces arising to avoid the flux flow (or whatever) dissipation mechanism act
against the exit of $\bf J$ from a region $\region$ with a very high slope. However, the evolution 
within $\region$ is that of a perfect conductor. In the limit of an infinite barrier,
the reaction is {\bf perpendicular} to the boundary of $\region$ (denoted by $\partial\region$). Thus, starting from an initial configuration with ${\bf J}\in \region$, and under the action of a transient electric field, the vector $\bf J$ quickly touches and/or shifts along the boundary, until a point is reached where the condition ${\bf E}\perp \partial \region$ is fulfilled. Owing to the perfect conductivity condition $\partial_{t}{\bf J}\propto{\bf E}$ no further evolution can occur (see Fig. \ref{fig_3} and Ref.\onlinecite{normal}). Faraday's law will be eventually the key for determining the actual point.
\end{itemize}

Recall that the above rule can be expressed as the condition of maximal
${\bf J}.\hat {\bf E}$ projection for ${\bf J }\in \region$. Recall also that the fundamental property already discussed for 1--D systems is verified: still the modulus $|{\bf E}|$ is irrelevant. On the other hand, notice that the {\em sign rule} is nothing but the 1--D particular case of the general {\em maximum projection rule}. The mathematical consistency is therefore satisfied. Let us now discuss on the physical soundness of the theory. This aspect is apparently related to the selection of the region $\region$ appropriate for the problem under consideration.

%
%

The simplest assumption that
translates the CS idea to 3--D situations was already issued by Bean
in Ref. \onlinecite{beanjap}. It has been called the {\em isotropic
CS model} and generalizes 1--D Bean's law to
\begin{equation}
\label{eq:cs3D}
{\bf J} =J_{c}\,\hat {\bf E}\quad{\rm if} \quad E \neq 0 \, ,
\end{equation}
i.e., the region $\region$ becomes a sphere.
This model has been used by several
authors; in spite of its mathematical simplicity, a remarkable
predictive power for reproducing a number of experiments with rotating and
crossed magnetic fields has been noticed,\cite{badiaPRL,badiaPRBvector,papersiso,baltaga} at least
qualitatively. It lacks, however, a solid physical basis. In any case, one could argue that statistical averaging over a system of entangled flux lines within a random pinning structure might be responsible for the isotropization of $\region$.

As stated above, to the moment, the most general theory for CS problems, formulated
in terms of a well accepted physical basis is the so called {\em
double critical state model}\,\cite{dcsm} (DCSM). In brief, this theory
assumes two different critical parameters, $J_{c\parallel}$ and $J_{c\perp}$
(see Fig. \ref{fig_3}) acting as the thresholds for the components of ${\bf J}$ parallel and perpendicular to ${\bf B}$ respectively. As stated above, $J_{c\perp}$ relates to
the flux depinning threshold induced by the Lorentz force on flux
tubes, while the additional $J_{c\parallel}$ is imposed by a
maximum gradient in the angle between adjacent vortices before mutual cutting and
recombination occurs.\cite{dcsm} The DCSM may be expressed by the statement
\begin{eqnarray}
\label{eq:dcsm}
\left\{
\begin{array}{ll}
{\bf J}_{\parallel} & =J_{c\parallel}\;\hat {\bf E}_{\parallel}\quad{\rm if} \quad {\bf E}_{\parallel} \neq 0\\
{\bf J}_{\perp} & =J_{c\perp}\,\hat {\bf E}_{\perp}\quad\!\!{\rm if} \quad {\bf E}_{\perp} \neq 0
\end{array}
\right.
\end{eqnarray}
Within the DCSM, the region $\region$ is a cylinder with its axis parallel to $\bf B$,
and a rectangular
longitudinal section in the plane defined by the unit vectors
$\hat{\bf B},\hat{\bf J}_{\perp}$ (see
Fig. \ref{fig_4}). The edges of the region $\region$ introduce a criterion
for classifying the CS configurations into : (i) T-states where the
flux depinning threshold has been reached (${\bf J}$ belongs to the
horizontal sides of the rectangle), (ii) C-states where the cutting
threshold has been reached (${\bf J}$ belongs to the vertical sides
of the rectangle), and (iii) CT-states where both $J_{\parallel}$
and $J_{\perp}$ have reached their critical values (corners of the
rectangle). 

%
%

Notice that $J_{c\parallel}$ and
$J_{c\perp}$ are determined from different physical phenomena, and their values may be very different (in general
$J_{c\parallel}>J_{c\perp}$ or even $J_{c\parallel}\gg J_{c\perp}$). Nevertheless, the
coupling of parallel and perpendicular effects is suggested by
experiments\cite{boyer80} and, for instance, may be included in the
theory by the condition $J_{c\parallel}=K B J_{c\perp}$ with $K$ a
material dependent constant. Recalling that the mesoscopic parameters $J_{c}$ are related to averages over the flux line lattice, interacting activation barriers for the mechanisms of flux depinning and cutting are expected and this may give place to deformations in the boundary $\partial\region$. Then, the theory should be able to host different regions as the ones depicted in Fig. \ref{fig_4}, or the situation suggested in Ref. \onlinecite{brandt_unusual} (Fig. 1). Regarding that proposal, we want to emphasize that the statement qualifying our maximum projection rule (i.e.: max\; ${\bf J}\cdot{\bf \hat{E}} \Leftrightarrow {\bf E}\perp\partial\region$) as physically incorrect in some cases, because the direction of ${\bf E}$ should be different, must be reconsidered. In fact, as exposed above, in the CS theory the material law allows some ambiguity described by an arbitrary scalar function in the form of an additional
potential term, i.e.: the rule ${\bf E} = {\bf E}_{\rm cs} + \btd \Phi$ does not affect the magnetic variables.
Here, ${\bf E}_{\rm cs}$ represents the electric field obtained from the maximal shielding rule,
and ${\bf E}$ a possible modification in order to adjust the electrostatic sector,
f.i. the scalar condition ${\bf E}\cdot{\bf B} = 0$ (${\bf E}\perp{\bf B}$) for the T-states. Thus, the maximal shielding rule can be easily 
complemented with an additional equation $\btd \Phi  \cdot {\bf B} =
- {\bf E}_{\rm cs} \cdot {\bf B}$ for the new variable $\Phi$ that allows to link our theory to many other models, expressed by an ${\bf E}({\bf J})$ law. Eventually, not only
the magnetic sector but also the electrostatic one would coincide in both formulations. In physical terms, one can state the theory as follows. First, the magnetic response is described by the Critical State law, that determines the relative orientation of ${\bf E}_{\rm CS}$ and ${\bf J}$ through the maximum projection rule. If required, the orientation of the full electric field may be tuned by considering the electrostatic charges.

%
\subsection{The variational principle: a general treatment}
\label{varprin}

The CS theory has been formulated as a minimization principle by several groups (see f.i.: Refs.\onlinecite{bossavit,prigozhin,badiaPRL,badiaPRBvector,sanchez}). In our case,\cite{badiaPRL,badiaPRBvector} an Optimal Control\cite{pontryagin} variational statement was introduced
for dealing with general CS problems. In previous work,\cite{badiaJLTP} a number of experimental facts were discussed in terms of different choices for the {\em control region} $\region$. Here, we want to emphasize that the variational principle is
not to be associated to any particular model, i.e., an arbitrary selection of the region $\region$ is allowed. 

Below, we summarize the main features of the formulation.
Its full equivalence to the approach based on the differential
equation statements will be shown. The hypotheses about constitutive
relations and the sweep rate independence made in the earlier
analysis will be obviously maintained. The principle is based on a
discretization of the path followed by the external sources, that
is, it is an approximation to the continuous evolution whose
accuracy increases as the step diminishes. 

Let us consider a small path step, from some initial profile of the magnetic field ${\bf B}_{\rm n}({\bf r})$ to a final profile ${\bf B}_{\rm n+1}({\bf r})$,
(define $\Inc{\bf B} = {\bf B}_{\rm n+1} - {\bf B}_{\rm n}$, and also the corresponding ${\bf J}_{\rm n}({\bf r})$ and ${\bf J}_{\rm n+1}({\bf r})$). Both configurations can be considered to be connected by a stationary process, i.e., we perform a small linear step $\Inc{\bf B}$, such that ${\bf B}_{\rm n+1} = {\bf B}_{\rm n} + s\Inc{\bf B}$,
$s\in [0,1]$. The initial condition fulfills Ampere's law $\btd \times {\bf B}_{\rm n} = \mu _0
{\bf J}_{\rm n}$, as well as $\btd \cdot{\bf B}_{\rm n} = 0$, $\btd \cdot {\bf J}_{\rm n} = 0$. As shown in Ref. \onlinecite{badiaPRBvector}, maximal shielding can be implemented by imposing the minimization of the step variation for the magnetic field profile integral, i.e.:
\begin{equation}
\label{eq:minprin}
{\cal F}[{\bf B}_{\rm n+1}(\cdot)] \equiv {\rm Min} \int _{\IR ^3} \,\frac {1}{2}(\Inc {\bf B}) ^2 d^3{\bf r}
\,\,{\rm .}
\end{equation}

Recall that minimization must be performed under the restrictions on the final profile: (i) Ampere's law $\btd \times {\bf B}_{\rm n+1}
= \mu _0 {\bf J}_{\rm n+1}$, and (ii) ${\bf J}_{\rm n+1}({\bf r}) \in \region$. This is a minimization problem within the variational calculus framework 
(integral functionals of unknown fields and their derivatives) with constraints,  
that can be analyzed with the tools of the Optimal Control theory.\cite{pontryagin}

Following
the usual Lagrange multipliers method, we build a ``Lagrangian''

\begin{equation}
L\equiv  \frac {1}{2}(\Inc{\bf B})^2 + {\bf p}\cdot(\btd \times
{\bf B}_{\rm n+1} - {\bf J}_{\rm n+1})
\end{equation}
that enforces Ampere's law. In fact, the Euler--Lagrange equations become
\begin{equation}
\label{eq:ampere}
\btd \times {\bf B}_{\rm n+1} - {\bf J}_{\rm n+1} = 0 \,
\end{equation}
for arbitrary variations $\delta {\bf p}$ of the multipliers, and
\begin{equation}
\label{eq:faraday}
\btd \times {\bf p} = - \Inc{\bf B}\, ,
\end{equation}
for arbitrary variations $\delta({\bf B}_{\rm n+1})$.

The second condition identifies $\bf p$ with $- \Inc{\bf
A}$ (recall that $\btd \times \Inc{\bf A} = \Inc
{\bf B}$). Then, one gets the {\em critical state} electric field ${\bf E}_{\rm cs} \Delta t= - \Delta{\bf A} =
{\bf p}$. Concerning the  ``parameter'' ${\bf J}_{\rm n+1}$, as far
as it is not allowed to take arbitrary values, we cannot impose
arbitrary variations as it is customary for the typical stationarity condition of the
Euler--Lagrange equations. Instead, an Optimal Control--like Maximum
principle must be used.\cite{badiaPRL} The minimum of the Lagrangian
must be sought within the set of current density vectors fulfilling
${\bf J}\in \region$, i.e.: ${\bf J}_{\rm n+1}$ is determined by the condition
\begin{equation}
\label{eq:max}
{\rm Min}\{L\}\,\left|_{{\bf J}\in \region}\right.\quad\equiv\quad{\rm Max} \;\{{\bf J}\cdot{\bf p}\}\,\left|_{{\bf J}\in \region}\right. \, .
\end{equation}
Notice that the maximal shielding condition is equivalent to the maximum projection rule, i.e.: the orthogonality condition
of the electric field direction with the surface of $\region$
previously discussed (Sec.\ref{matlaw}) is recovered. Notice also that Ampere's law is imposed
[Eq. (\ref{eq:ampere})] through the Lagrange multiplier, while the discretized version of Faraday's
law [Eq. (\ref{eq:faraday})] is derived as an Euler--Lagrange equation for the variational problem, so that absolute consistency
with the Maxwell equations is obtained. Moreover, maximal global
(integral) shielding is achieved through a maximal local
 shielding rule [Eq. \ref{eq:max}] that reproduces the elementary evolution of
$\partial_{t}{\bf J}$ for a perfect conductor with restricted currents.

In 3--D problems, as an advantage of the formulation in Eq. (\ref{eq:minprin}), one can avoid the integration of the equivalent partial differential equations and straightforwardly minimize the discretized integral by using a numerical algorithm for constrained minimization. It is this numerical minimization, instead of numerical integration of PDEs, which represents a very important advantage
in the performance and power of the variational method. The Lagrange multiplier $\bf p$ (basically, the electric field)
disappears in the direct minimization process, while
the magnetic field can be expressed in terms of an external contribution $\mu _0 {\bf H}_{0}$ and the local sample's currents. As a consequence, only the unknown current components appear
in the computation, reducing the number of unknown variables. Any symmetry of the problem will allow further simplifications and correspondingly faster numerical convergence. 

Being more specific, the integrand $\frac {1}{2} (\Inc{\bf B}) ^2$
can be rewritten as $\frac {1}{2} (\Inc{\bf B})\cdot (\btd \times \Inc{\bf A})$, and manipulated to get $\frac {1}{2} (\Inc{\bf A})\cdot (\btd \times \Inc{\bf B})$ plus a divergence term, fixed by the external sources at a distant surface. Now, the integral is restricted to the samples region $\Omega$, because
$\btd \times \Inc{\bf B} = \mu _0 \Inc{\bf J}$ is only unknown within the superconductor. In addition, the vector potential can be expressed as 

\begin{equation}
 \Inc{\bf A} = \Inc{\bf A}_{0}
+ \frac {\mu _0 }{4 \pi} \int _\Omega \frac {\Inc{\bf J}}{|{\bf r} - {\bf r}'|} d^3 {\bf r}' \, .
\end{equation}

This transforms ${\cal F}$ into a double integral over the body of the sample, i.e.:

\begin{eqnarray}
\label{eq:genMij}
&&{\cal F} = \frac {8\pi}{\mu_0}\int _\Omega \Inc{\bf A}_{0}\cdot {\bf J}_{\rm n+1}({\bf r})d^3 {\bf r}
\nonumber\\
&&+\int\!\int _{\Omega \times \Omega}
\frac {{\bf J}_{\rm n+1}({\bf r}')
\cdot [{\bf J}_{\rm n+1}({\bf r}) - 2 {\bf J}_{\rm n}({\bf r})]}{|{\bf r} - {\bf r}'|}d^3 {\bf r}d^3 {\bf r}'\;
\end{eqnarray}

(terms independent of ${\bf J}_{\rm n+1}$ have been omitted). 

Finally, in addition to the incorporation of the external sources (${\bf A}_{0}$), and the constraints for the allowed {\bf J} region $\region$ one must also ensure the charge conservation condition by searching the minimum for the allowed set of current densities fulfilling $\nabla\cdot{\bf J}=0$ as an additional constraint.

%
%

\section{General critical states: applications}
\label{application}

In this section, we will show that the variational statement may be
used to predict the magnetic structure for the {T-states} in a {\em
three-dimensional} slab geometry, i.e.: both in-plane and
perpendicular magnetic field components are applied to an infinite
slab and varied in a given fashion. A wide range of applied fields will be considered, and our results compared to those available in the literature. Moreover, we will study the
corrections that appear when a more general {\em CT-state}
framework is introduced.

First, we will give the details related to the mathematical statement of
the general critical state in the slab geometry. Then, the theory
will be applied for establishing the capabilities of several
versions of the DCSM within different physical scenarios.

\subsection{Infinite slab: general double critical state}
\label{TandCT}

Within the infinite slab geometry, the variational formulation of
the general DCSM allows an algebraic statement that is rather
convenient for the eventual numerical application. To be specific,
we will consider an infinite slab, that is fully penetrated by a
perpendicular uniform field $H_{z0}$ and then subjected to a certain
process for the applied parallel field, i.e.:
$[H_{x0}(t),H_{y0}(t)]$ as indicated in Fig. \ref{fig_5}. Recalling the symmetry
properties of the electromagnetic quantities, one can describe the
problem as a stack of current layers parallel to the sample's
surface. Assuming that the slab occupies the space $|z|\leq d/2$, it
suffices to discretize the upper half, i.e.: $0\leq z \leq d/2$ as
symmetry (or antisymmetry) conditions may be applied. Thus, in what
follows, a collection of $N$ layers ($z_{i}=\delta\, i\;,\;
\delta\equiv d/2N$) will be considered. Within this approximation,
one has to include two components of ${\bf J}$ within each layer,
i.e.: $[J_{x}(z_{i}),J_{y}(z_{i})]$. Notice that position
independence for a given value of $z$ ensures a divergenceless
${\bf J}$. Furthermore, a sheet current may be introduced. Thus, the
problem will be stated in terms of $\xi_{i}\equiv
J_{x}(z_{i})\delta$ and $\psi_{i}\equiv J_{y}(z_{i})\delta$
($\delta$ denoting the width of the layers). Now, a straightforward
application of Amp\`ere's law allows to express the penetrating
magnetic field as the sums over the layers

{
\begin{eqnarray}
\label{eq:fields}
H_{x}(z_i)\equiv H_{x,i}&=&-\sum_{j>i}\psi_{j}-\psi_{i}/2
\nonumber\\
H_{y}(z_i)\equiv H_{y,i}&=&\sum_{j>i}\xi_{j}+\xi_{i}/2 \, .
\end{eqnarray}
}

Next, we recall that in the slab geometry Eq. (\ref{eq:minprin}) becomes a discretized principle
restricted to the volume of the slab. Following the concept introduced in the previous section (see Eq. \ref{eq:genMij}), the problem may be transformed into the minimization over the current densities

\begin{eqnarray}
\label{eq:minprindis} {\tt min\;\;F}&=&\displaystyle{\displaystyle
\frac{1}{2}}\sum_{i,j}\xi_{i,{\rm n+1}}M_{ij}^{x}\,\xi_{j,{\rm n+1}}
-\sum_{i,j}\xi_{i,{\rm n}}M_{ij}^{x}\,\xi_{j,{\rm n+1}}
\nonumber\\
&+&\displaystyle{\displaystyle \frac{1}{2}}\sum_{i,j}\psi_{i,{\rm
n+1}}M_{ij}^{y}\,\psi_{j,{\rm n+1}} -\sum_{i,j}\psi_{i,{\rm
n}}M_{ij}^{y}\,\psi_{j,{\rm n+1}}
\nonumber\\
&-&\sum_{i}\psi_{i,{\rm n+1}}(i-1/2)(H_{x0,{\rm n+1}}-H_{x0,{\rm n}})
\nonumber\\
&+&\sum_{i}\xi_{i,{\rm n+1}}(i-1/2)(H_{y0,{\rm n+1}}-H_{y0,{\rm
n}}) \, .
\end{eqnarray}

We stress that minimization has to be performed under the restrictions $J_{\parallel}\leq J_{c\parallel}$ and $J_{\perp}\leq J_{c\perp}$, i.e.: the DCSM hypotheses. Specifically, one has to invoke the conditions

\begin{eqnarray}
\label{eq:jpape}
(1-h_{x,i}^{2})\xi_{i}^{2}+(1-h_{y,i}^{2})\psi_{i}^{2}-2h_{x,i}h_{y,i}\,\xi_{i}\psi_{i}\leq
j_{c\perp}^{2}
\nonumber\\
\nonumber\\
h_{x,i}^{2}\,\xi_{i}^{2}+h_{y,i}^{2}\,\psi_{i}^{2}+2h_{x,i}h_{y,i}\,\xi_{i}\psi_{i}\leq
j_{c\parallel}^{2} \, . 
\end{eqnarray}
(Here, the normalization ${\bf h}\equiv{\bf H}/H$ has been used.)

Recall that the subindex $n$ is introduced to indicate
time discretization, i.e.: $\psi_{i}(t+\Delta t)-\psi_{i}(t)\equiv
\psi_{i,{\rm n+1}}-\psi_{i,{\rm n}}$. When this index is omitted,
it will be meant that the equation is time independent (it is valid $\forall\;n$ actually).

Finally, the reader can check that a straightforward substitution of the squared
components of the magnetic field entering the expression in Eq. (\ref{eq:minprin}) in terms of Eq. (\ref{eq:fields})
leads to the following formulas for the mutual inductance
coupling between layers

\begin{eqnarray}
\label{eq:mcoef} M_{ij}^{x}&=&M_{ij}^{y}\equiv 1+2\left[{\rm
min}\left\{ i,j\right\}\right]\quad \forall\; i\neq j
\nonumber\\
M_{ii}^{x}&=&M_{ii}^{y}\equiv 2\left(\frac{1}{4}+i-1\right)
\end{eqnarray}

Below, we present a number of results obtained by application of the
previous equations for the slab. First, an infinite band model ($J_{\parallel}\ll J_{c\parallel}$ or the so-called T-states) will be
considered. Afterward, the corrections related to the flux cutting
limitation ($J_{c\parallel}$) will be studied. When possible, our
results will be compared with available literature.

%
%
\subsubsection{T-states in 3--D configurations}\label{T-states3D}

%
%
\begin{figure}[!]
\begin{center}
{\includegraphics[width=0.48\textwidth]{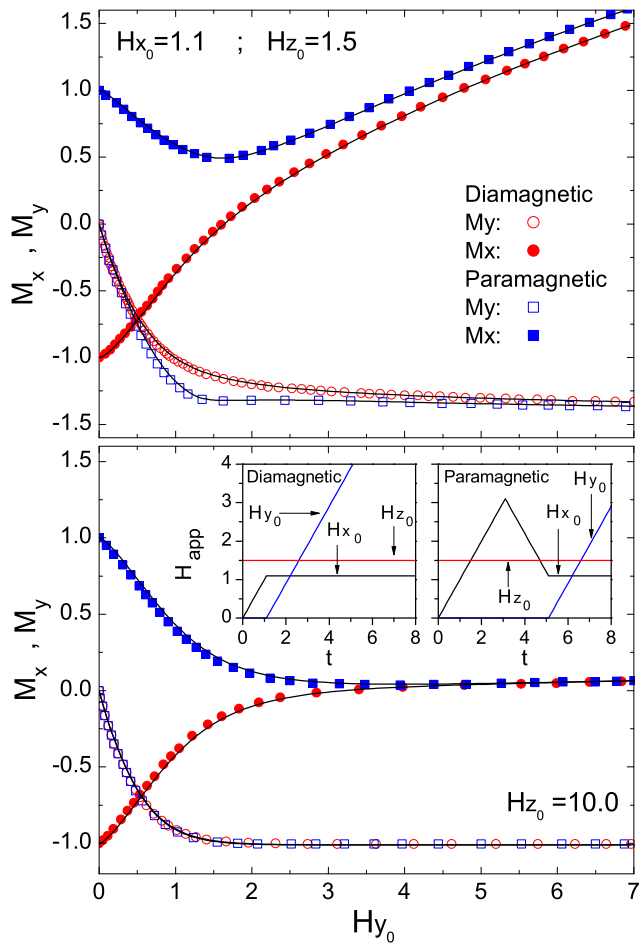}}
\caption{\label{fig_5}(color online) The magnetic moment
($M_{x},M_{y}$) of the slab defined by Eq. (\ref{eq:M})
as a function of $H_{y0}$. Shown are the diamagnetic and
paramagnetic cases for $H_{x0}=1.1$, and $H_{z0}=1.5$ (top) and
$H_{z0}=10$ (bottom). The experimental processes for
the applied magnetic fields are shown as insets in the bottom panel.
Units are $j_{c\bot}d/2$ for $H$ and $j_{c\bot}d^{2}/4$ for $M$. Our
results (lines) are compared to those by Brandt and Mikitik in
Ref. \onlinecite{brandt_unusual} (symbols). In all these cases, we have taken $J_{c\parallel}=\infty$
and $J_{c\perp}=1$ (T-states).}
\end{center}
\end{figure}

Here, we show the theoretical predictions for the T-states along the
magnetization process indicated in the insets of Fig. \ref{fig_5}.
Starting from a fully penetrated state, with a field applied
perpendicular to the slab surface ($H_{z0}$), one configures either
a diamagnetic or a paramagnetic critical state by sweeping the
applied parallel component $H_{x0}$ (thus inducing $J_y$). Eventually, an
increasing ramp in the other field component, $H_{y0}$ (thus inducing
$J_x$) is applied. The response of the superconductor is obtained as
a collection of values for the sheet currents $\{\xi_{i},\psi_{i}\}$
at the forward time layers $n=1,2,3,\dots$. The magnetic field
profiles and magnetic moments are eventually obtained by numerical
integration. In particular, the expression of the magnetic moment per unit area

\begin{equation}
\label{eq:M}
{\bf M}=\frac{1}{2L}\int_{\rm Vol}\; {\bf z}\times{\bf J}d^{3}{\bf r}
=\int_{-d/2}^{d/2}\; {\bf z}\times{\bf J}dz
\end{equation}

has been used, with L representing the length of the sample. Recall that we have invoked the property that for long loops, the contribution coming from the U-turn at the ends, exactly equals the contribution of the long sides. This may be shown starting from the condition ${\nabla}\cdot{\bf J}=0$ (no sources) that allows to consider the current density distribution as a collection of loops and ensures the equality of the integrals over $zJ_y$ and $zJ_x$ (see Ref.~ \onlinecite{brandt94}).

Owing to the rich phenomenology encountered, the results will be given separately for moderate and low perpendicular fields. Recalling that $H$ is measured in units of the physically relevant penetration
field $J_{c\perp}d/2$, then $H_{z0}=0.1, 1.5, 10$ will cover the range of interest.

%
%
\begin{figure}
\begin{center}
\includegraphics[width=0.48\textwidth]{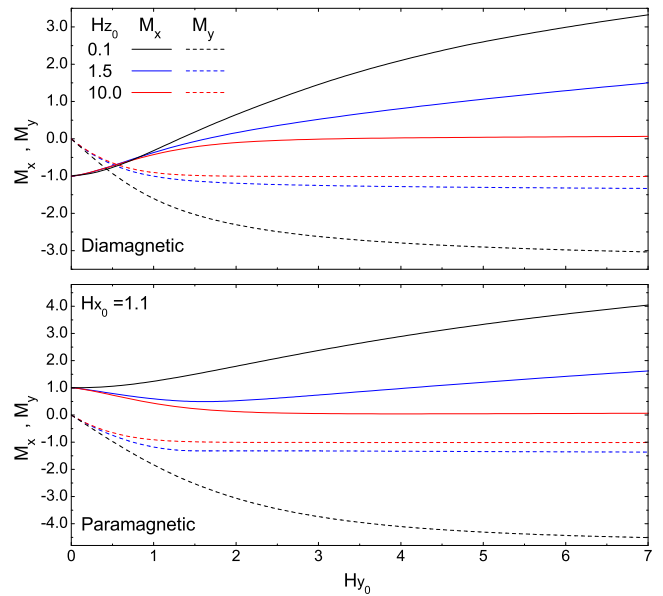}
\caption{\label{fig_6} (color online) The magnetic moment $M_{x}$
(solid lines) and $M_{y}$ (dashed lines) for the T-states, as a function of $H_{y0}$.
Shown are the diamagnetic (top) and paramagnetic (bottom) cases for
$H_{z0}=0.1$ (black), $H_{z0}=1.5$ (blue), and
$H_{z0}=10.0$ (red). Units are $j_{c\bot}d/2$ for $H$ and
$j_{c\bot}d^{2}/4$ for $M$.}
\end{center}
\end{figure}

\paragraph{Moderate fields} Fig. \ref{fig_5} shows our results for $H_{z0}=1.5$ and $H_{z0}=10$
compared to those of Ref. \onlinecite{brandt_unusual} obtained under
the same conditions. A remarkable agreement is to be noticed, thus
validating our theory against the differential equation approach of
that paper. The same degree of coincidence was also checked for the magnetic field and current densities (not shown for brevity) as expected. In fact, the equivalence of our maximum projection rule and the ${\bf E}({\bf J})$ law based analysis\cite{brandt_unusual} may be proofed as follows. The material law in that work was applied in two steps: (i) the transient electric field was
chosen along the direction ${\bf B}\times({\bf
J}\times{\bf B})$ as dictated by the flux flow condition, and  (ii) the
magnitude of ${\bf E}$ was found from the condition
$J_\perp=J_{c\perp}$, through the Amp\`ere and Faraday's laws and appropriate boundary conditions.
We recall that $J_\perp=J_{c\perp}$ is equivalent to the selection of our horizontal band for $\region$, that the direction of $\hat{\bf E}$ is straightforwardly the same, and finally that the remaining component $J_{\parallel}$ is also coincident as it is obtained from the Maxwell equations, also contained in the variational formulation.

%
%
\begin{figure}[t]
\begin{center}
\includegraphics[width=0.47\textwidth]{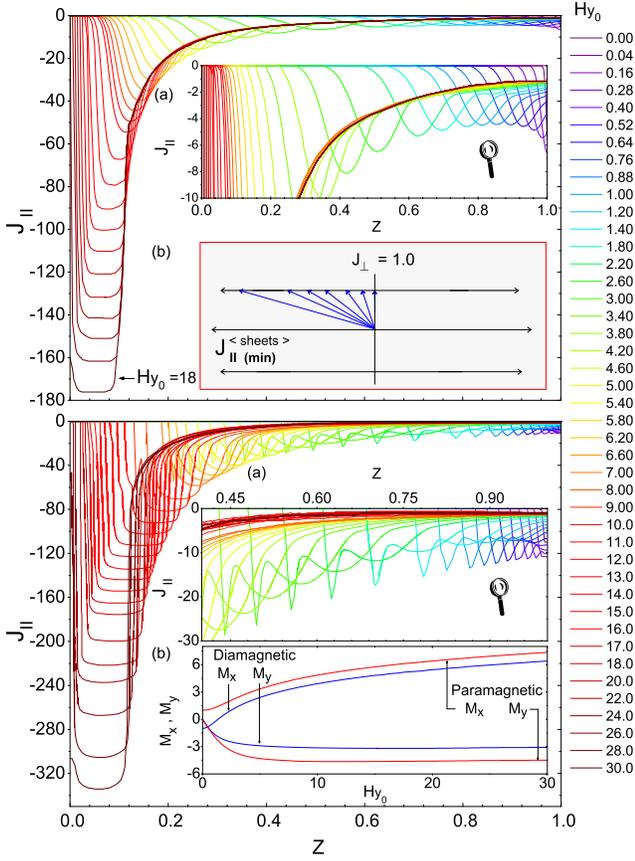}
\caption{\label{fig_7} (color online) Profiles of the component
$J_{\parallel}$ for the limit $J_{c_{\parallel}}\to\infty$ (T-state) with
$H_{z0}=0.1$. In all cases the perpendicular current profiles satisfy $J_{\bot}=J_{c_{\bot}}=1.0$. The diamagnetic (top) and paramagnetic
(bottom) cases are shown. Top: inset (a) shows a zoom of $J_{{||}}$ for the first profiles of $H_{y_{0}}$. Inset
(b) schematically shows the evolution of the vector $\bf J$ as
function of its parallel and perpendicular components. Bottom: inset (a) shows a zoom of $J_{{||}}$
for the first profiles of increasing $H_{y0}$. Inset (b) shows the magnetic
moment components ($M_{x}$, $M_{y}$) per unit area as a function of
$H_{y0}$.}
\end{center}
\end{figure}

%
%
\begin{figure}
\begin{center}
\includegraphics[width=.5\textwidth]{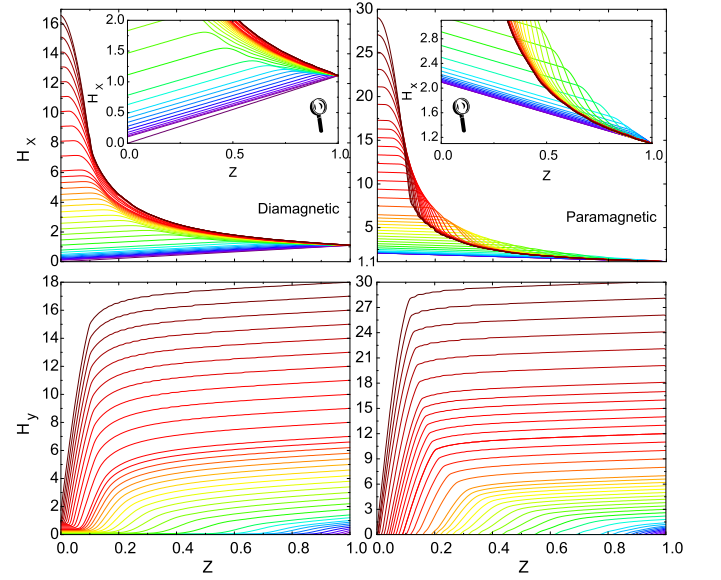}
\caption{\label{fig_8} (color online) Magnetic field components
$H_{x}(z)$ (top) and $H_{y}(z)$ (bottom) corresponding to the current density profiles
for the T-state limit with
$H_{z_{0}}=0.1$ (Fig. \ref{fig_7}). The diamagnetic (left) and
paramagnetic (right) cases are shown. The insets shown a zoom of
the corresponding pictures.}
\end{center}
\end{figure}

\paragraph{Low fields} In Fig. \ref{fig_6} we display the effect of extending the previous results to the low field region ($H_{z0}=0.1$), by comparison to the values $H_{z0}=1.5$ and $H_{z0}=10$. The plots indicate the following features. (i) in general,
a saturation is reached for $M_{y}(H_{y0})$, as compared to the
eventual linear increase of $M_{x}(H_{y0})$ for the highest values
of $H_{y0}$, (ii) the higher $H_{z0}$, the sooner the saturation is
reached, (iii) increasing $H_{z0}$ rapidly diminishes the slope of
$M_{x}(H_{y0})$, (iv) in the paramagnetic case, a minimum is
observed (more evidently for $M_x$, and more visible in Fig. \ref{fig_5} for moderate $H_{z0}$, that is
smoothed either for the higher or lower values of this field
component. 

As indicated above, the underlying flux penetration profiles for the moderate field region were already presented in Ref. \onlinecite{brandt_unusual} and fully coincide with our calculations. However, the low field region was uncovered. Here, we will show the peculiar behavior of the field and current density profiles for this regime. Thus, Fig. \ref{fig_7} displays the behavior of the projection of the current density onto the direction of the magnetic field ($J_{\parallel}$) under the ansatz of a T-state structure for $H_{z0}=0.1$. Recall that, hereafter, $z$ is given in units of $d/2$. Then, $z=0$ corresponds to the center of the sample and $z=1$ to the surface.

It is apparent that the full penetration of the T-state requires a high field component ($H_{y0}\approx  18$ and $H_{y0}\approx  30$ for the dia- and paramagnetic cases respectively), and a very high ratio $J_{\parallel}/J_{c\perp}$ ($\approx 180$ for the diamagnetic case and $\approx 340$ for the paramagnetic one). Notice that until these values are reached, one has $J_{\parallel}=0, J_{\perp}=1$ for the inner part of the sample, and a certain distribution $J_{\parallel}(z)$ for the outer region. We also recall a somehow complex structure with one or two minima in between the surface of the sample and the point reached by the perturbation. Interestingly, when $H_{y0}$ grows, the minima become very flat, corresponding to a nearly constant value of $J_{\parallel}$. From the physical point of view, the minima basically represent the region where ${\bf H}$ rotates so as to accommodate the penetration profile ${\bf H}(z)$ to the previous state of magnetization $(H_{x},0,H_{z0})$. From the point of view of Faraday's law, this takes place as quickly as possible so as to minimize flux variations. The obtained magnetic field profiles are shown in Fig. \ref{fig_8}. Their interpretation, in terms of the critical current restrictions (T-states), is simplified by the increasing value of $H_{y0}$. Thus, a steep variation of $H_y$ occurs for the inner region of the sample, corresponding to the large values of $J_{\parallel}$ (essentially $J_x$ because of the increasing $H_x$ in that region). On the other hand, $H_x$ displays a small slope, which relates to the condition $J_{\perp}=1$ (essentially, $J_{\perp}\approx J_{y}$ in the inner region).

%
%
%
%
\begin{figure}[!]
\begin{center}
\includegraphics[width=0.48\textwidth]{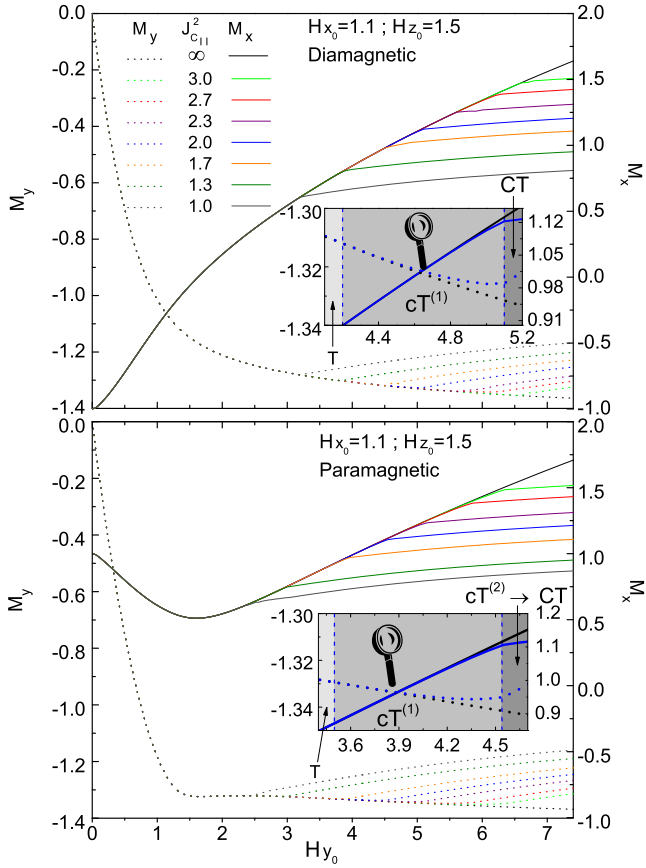}
\caption{\label{fig_9} (color online) The magnetic moment
($M_{x},M_{y}$) of the slab per unit area as a function of
$H_{y0}$. Shown are the diamagnetic (top) and paramagnetic (bottom)
cases for $H_{x0}=1.1$ and $H_{z0}=1.5$. The experimental conditions and units are those defined in Fig. \ref{fig_5}. The T-state curves ($J_{c\parallel}\gg 1$) are shown for comparison with several
rectangular cases: $J_{c_{||}}^{2}=$3.0, 2.7, 2.3, 2.0, 1,7, 1.3,
1.0. The insets show the particular case $J_{c_{||}}^{2}=2.0$ in
the region where the transition T$\rightarrow$CT is visible.}
\end{center}
\end{figure}

\subsubsection{CT-states in 3--D configurations}

In this section, we concentrate on the effect of considering a flux
cutting limitation ($J_{c\parallel}$). Magnetization curves, current
density and field penetration profiles will be shown, corresponding
to the same magnetic processes indicated in Fig. \ref{fig_5}, but now
for the rectangular DCSM regions with a number of values for $\chi =
J_{c\parallel}/J_{c\perp}$. In order to obtain continuity with the
T-state results (recall that, ideally this corresponds to the limit
$\chi\to\infty$) a range of increasing values for the parameter
$\chi$ will be analyzed. On the other hand, owing to the rich
phenomenology encountered, the results will be given separately for
moderate and low perpendicular fields, i.e.: $H_{z0}=1.5$ and $H_{z0}=0.1$.

\paragraph{Moderate fields}

\begin{figure}
\begin{center}
\includegraphics[width=0.48\textwidth]{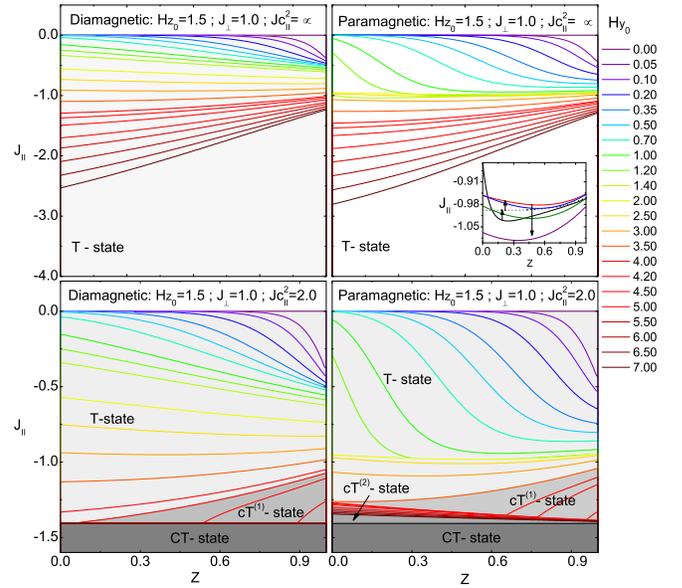}
\caption{\label{fig_10} (color online) Profiles of the parallel
currents $J_{||}$ for the T-state hypothesis ``$J_{c_{||}}\to\infty$ and
$J_{c_{\bot}}=1.0$'' (top) and for a rectangular DCSM with
``$J_{c_{||}}^{2}=2.0$ and $J_{c_{\bot}}=1.0$'' (bottom). The
diamagnetic (left) and paramagnetic (right) cases are shown. In the
paramagnetic case, the profiles of $J_{\parallel}$
for $H_{y0}=1.35,\, 1.7,\,2.0,\, 2.3,\, 2.6,\, 3.0$ are shown as an inset, and correspond to the sign change in the slope of the magnetic moment $M_{x}$. See the text for the definitions of the states cT$^{(1)}$, cT$^{(2)}$, and CT.}
\end{center}
\end{figure}

\begin{figure}[h]
\begin{center}
\includegraphics[width=0.48\textwidth]{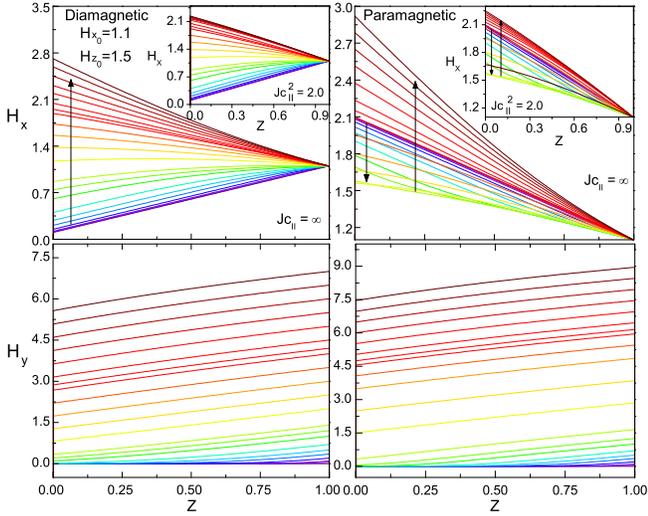}
\caption{\label{fig_11} (color online) The magnetic field components
$H_{x}(z)$ (top) and $H_{y}(z)$ (bottom) in the same critical states
of Fig. \ref{fig_10}. The diamagnetic (left) and paramagnetic
(right) cases are shown for the T-states and the indicated rectangular regions of ${\bf J}$. The profiles $H_{y}(z)$ show the
same behavior for $J_{c_{\parallel}}^{2}\to\infty$ (displayed) and $J_{c_{\parallel}}^{2}=2$(not shown).}
\end{center}
\end{figure}

The main facts for $H_{z0}=1.5$ are shown in
Figs. \ref{fig_9}-\ref{fig_11}. First, we plot the corrections to
$M_x$ and $M_y$ both for the dia- and para-magnetic cases, when the
DCSM region corresponds to the aspect ratio values $\chi^{2}=1.0,
1.3, 1.7. 2.0, 2.3, 2.7$ and $3.0$ (Fig. \ref{fig_9}). It is
noticeable that the limitation in $J_{c\parallel}$ produces a {\em
corner} in the magnetic moment dependencies $M_{x,y}(H_{y0})$, which
establishes the departure from the {\em master curve} defined by the
T-state. The corner in $M_x$ and $M_y$ appears at some characteristic
field $H_{y0}^{*}$ that increases with $\chi$, eventually
disappearing within the region of interest. The value $\chi^{*}$ for
which the corner is not observed, defines the range of application for
the T-state limit ($\chi^2 \gtrsim 3$ in the conditions of Fig. \ref{fig_9}). On the other hand, the fine structure of the corner is shown in the
insets of Fig. \ref{fig_9}. Notice that, in fact, the deviation from
the master curve takes place in two steps, being the second one that
really defines the corner.

%
%

In order to allow a physical interpretation on how the T-states
break down for the 3--D configurations studied in this section, in
Fig. \ref{fig_10} we plot the profiles of $J_{\parallel}(z)$ within
the slab, as $H_{y0}$ is increased. The upper panels show the
evolution of this quantity for the T-states, whereas the lower
panels show the process of saturation in which $J_{\parallel}$
reaches the value $J_{c\parallel}$ both for the dia- and
para-magnetic initial conditions. Just for convenience, we have
introduced the following notation. $cT$ denotes that $J_{\parallel}$
has reached the limit $J_{c\parallel}$ only partially within the sample, while $CT$
means that $J_{\parallel}$ equals $J_{c\parallel}$ for the whole
range $0\leq z\leq d/2$. For the {\em partial
penetration} cT-states, we additionally distinguish between the so-called
cT$^{(1)}$ and cT$^{(2)}$ phases. As one can see in the plot, cT$^{(1)}$
means that $J_{\parallel}$ penetrates linearly from the surface
until the limitation is reached somewhere within the sample. For the
diamagnetic case, the profile stops at the actual value
$J_{c\parallel}$. However, for the paramagnetic case, the structure
is more complex. Thus, $J_{\parallel}$ penetrates linearly until a
{\em linear increase} (toward the center) curve is reached. This structure is followed
until the contact between both lines reaches the surface. Then, the
so-called cT$^{(2)}$ region appears. $J_{\parallel}$ has reached
$J_{c\parallel}$ at the surface, and the whole $J_{\parallel}$ curve
``pivots'' around this point until the full CT-state is reached. We
call the readers' attention that the initial separations of the
magnetic moment from the T-state master curves take place as soon as a
cT$^{(1)}$-state is obtained. On the other hand, the corners can be clearly
assigned to the instant at which such a state disappears.

Just for completeness, the magnetic field penetration profiles,
corresponding to the $\chi^{2}\to\infty$ and $\chi^{2}= 2$ cases
are shown in Fig. \ref{fig_11}. Notice the change in curvature and
slope reduction in the penetration of $H_x$ for the CT states.
Notice also that the $H_y$ profiles are only shown for the T-states,
because a very similar behavior takes place (just differing in a
small compression for the higher values of $H_{y0}$).

%
%

\paragraph{Low fields}

%
%
\begin{figure}[t]
\begin{center}
\includegraphics[width=0.48\textwidth]{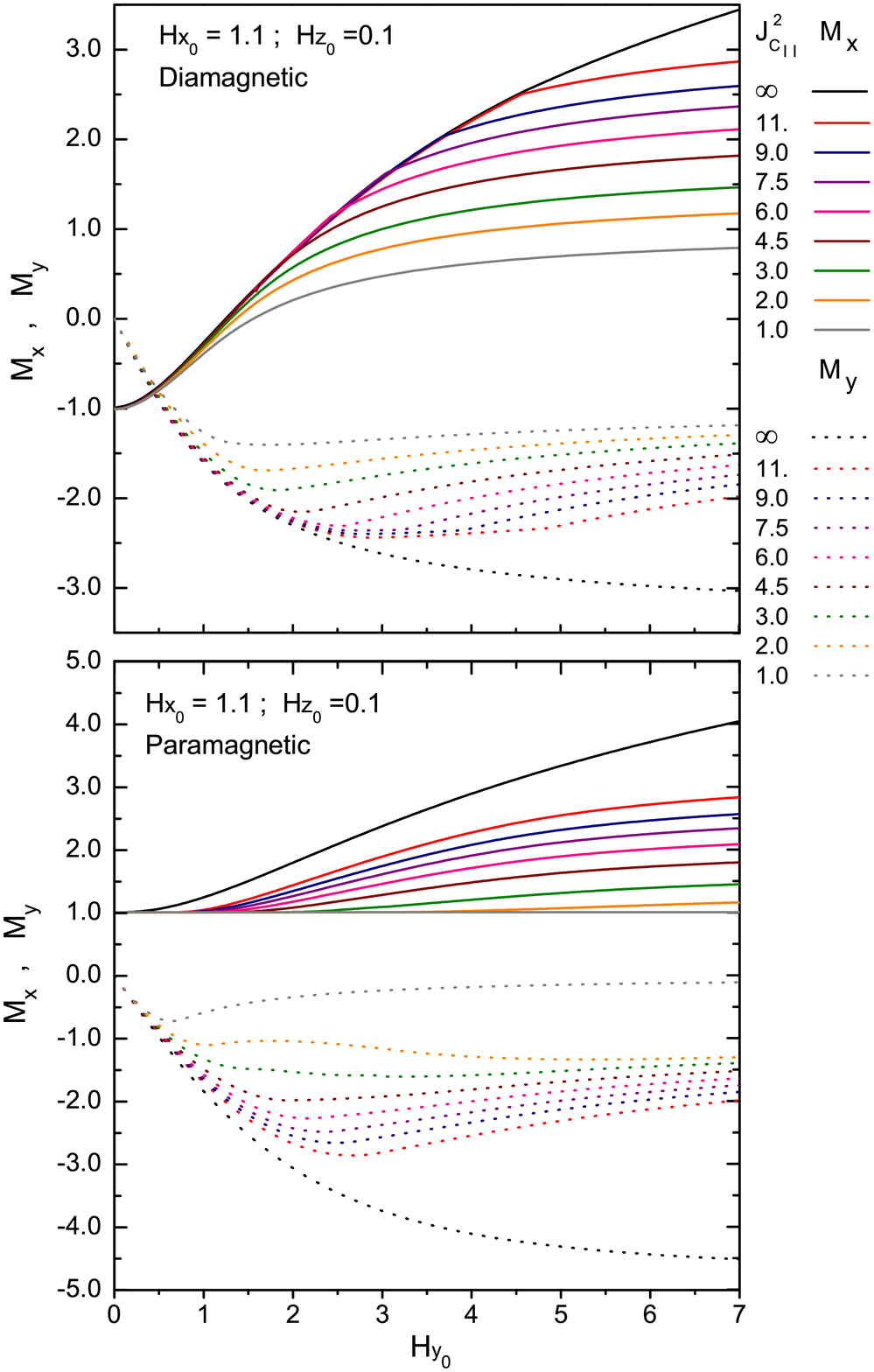}
\caption{\label{fig_12} (color online) Magnetic moment components
($M_{x},M_{y}$) of the slab as a function of
$H_{y0}$. Shown are the diamagnetic (top) and paramagnetic (bottom)
cases for $H_{x0}=1.1$ and $H_{z0}=0.1$. The infinite
band T-state model ($J_{c_{\parallel}}\to\infty$) is shown for comparison with several
rectangular cases: $J_{c_{\parallel}}^{2}=$ 11.0, 9.0, 7.5, 6.0, 4.5, 3.0,
2.0, and 1.0.}
\end{center}
\end{figure}

\begin{figure}[b]
\begin{center}
\includegraphics[width=0.48\textwidth, height=0.25\textwidth]{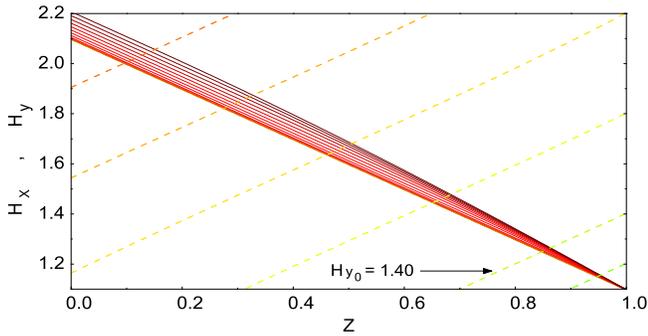}
\caption{\label{fig_13} (color online) Magnetic field components
$H_{x}(z)$ (solid-lines) and $H_{y}(z)$ (dashed-lines) corresponding
to the rectangular region $J_{c\parallel}^{2}=2.0$, $J_{c\perp}^{2}=1.0$ and initial paramagnetic conditions. The curves follow the color scale convention in Fig. \ref{fig_10}. For clarity, the $H_{y}(z)$ profile corresponding to $H_{y0}=1.40$ has been labeled accordingly. The analogous plot for the diamagnetic case strongly resembles that of Fig. \ref{fig_11}.}
\end{center}
\end{figure}

Although the general trends in the CT-state corrections for low $H_{z0}$ do not very much differ from those at moderate field values, some distinctive features are worth to be mentioned for the $M_{x,y}(H_{y0})$ curves. To start with, we recall that the corner structure that defines the separation of the {\em CT curves} from the {\em master T-state behavior} is different. Thus, as one can notice in Fig. \ref{fig_12}, it is only for the higher values of $\chi^{2}$ that the separations take place abruptly. In particular, a smooth variation occurs for $\chi^{2}<6$ in all cases. Also noticeable is the change in the behavior of the initial part of the $M_{x}(H_{y0})$ curves for the paramagnetic case. Recall that the minimum observed for the moderate field region ($H_{z0}=1.5$) has now disappeared (this can be already detected for the T-states). Significantly, what one can see as $\chi^{2}$ decreases is that $M_x$ develops a nearly flat region at the low values of $H_{y0}$. Physically, this means that the initial $H_{x}(z)$ profile is basically unchanged. For the lowest values of $\chi^2$ this can take place over a noticeable range of applied fields $H_{y0}$. A detail about the origin of this behavior can be seen in Fig. \ref{fig_13}, that corresponds to $\chi^{2}=2$. Notice the insignificant variation of $H_x$ as compared to the changes in $H_y$ along the process.

\begin{figure}
\begin{center}
\includegraphics[width=0.47\textwidth]{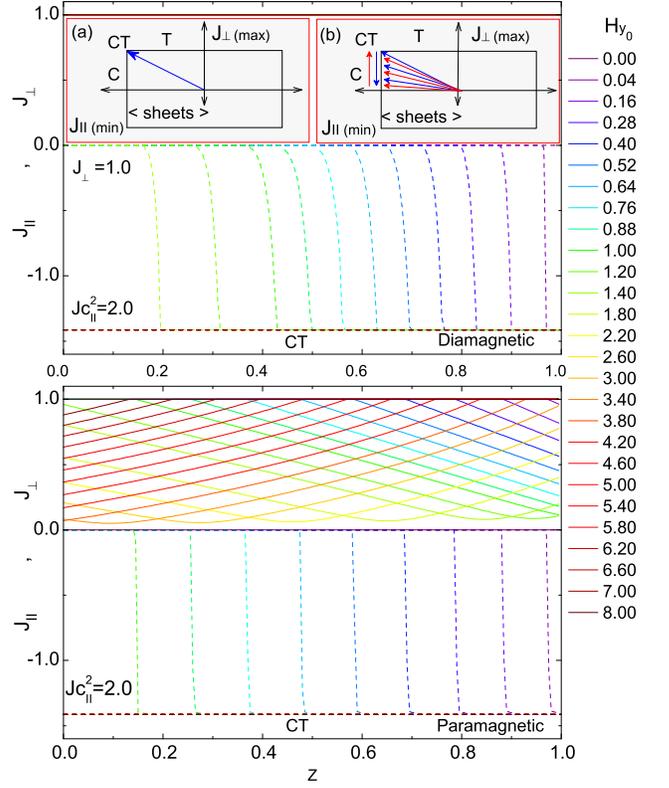}
\caption{\label{fig_14} (color online) Profiles of the parallel
($J_{||}$) and perpendicular ($J_{\bot}$) current densities for rectangular
region ``$J_{c_{||}}^{2}=2.0$, $J_{c_{\bot}}=1.0$'' with
$H_{z0}=0.1$. The diamagnetic (top) and paramagnetic (bottom)
cases are shown. Inset (a) schematically shows the CT structure of the full penetration regime in the diamagnetic case. The CT-C structure behavior of ${\bf J}$ for the paramagnetic case is shown in inset (b).}
\end{center}
\end{figure}

\begin{figure}
\begin{center}
\includegraphics[width=0.48\textwidth]{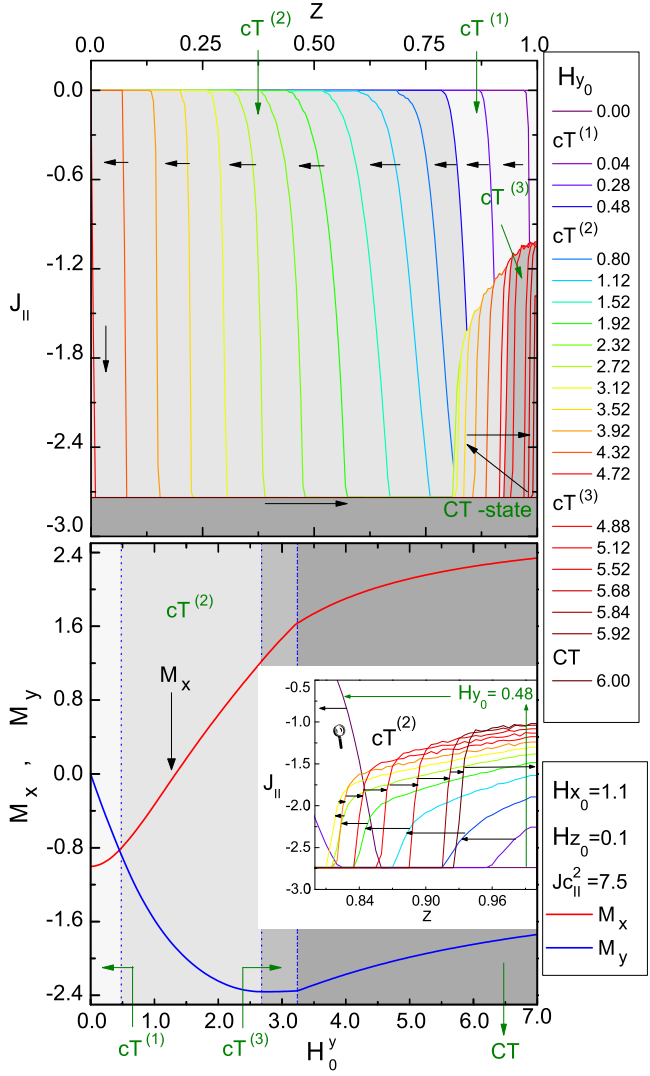}
\caption{\label{fig_15} (color online) Top: Profiles of $J_{||}$ for the diamagnetic case within the rectangular DCSM with $\chi^{2}=7.5$ and $H_{z0}=0.1$. In all cases, one gets 
$J_{\perp}=J_{c\perp}=1.0$. Bottom: The corresponding magnetic moment components
($M_{x},M_{y}$) as a function of $H_{y0}$ are shown. The evolution from the initial full penetration T state to the final full penetration CT state takes place in three steps that are classified according to the structure along the sample width by: cT$^{(1)}\equiv$  T-CT, cT$^{(2)}\equiv$ T-CT-T, cT$^{(3)}\equiv$ CT-T and eventually CT.}
\end{center}
\end{figure}

Also remarkable are the peculiarities of the current density penetration profiles for low values of $H_{z0}$. They can be observed in Figs. \ref{fig_14} and \ref{fig_15}, that reveal new physical mechanisms, that did not appear for the moderate perpendicular field values.
%
%
Again, the first observation is that the appearance of the corner in the magnetic moment straightforwardly relates to the current density profiles. Thus, for the lower values of $\chi$ (no corner present), the profile $J_{\parallel}$ displays a rather simple structure, basically jumping from $0$ to $J_{c\parallel}$ at some point within the sample (Fig. \ref{fig_14}). On the contrary, for the higher values of $\chi$ (those displaying a corner in $M(H_{y0}$) the evolution of the $J_{\parallel}(z)$ profiles is much more complex (Fig. \ref{fig_15}). Let us go into detail about these topics, part by part.
%
%
%

Fig. \ref{fig_14} shows the profiles $J_{\parallel}(z)$ both for the dia- and paramagnetic cases for $\chi^{2}=2$. Recall that the evolution of the profiles with the increase of $H_{y0}$ is very similar. The above mentioned step-like structure with $J_{\parallel}=0$ in the inner part and $J_{\parallel}=J_{c\parallel}$ in the periphery evolves until the {\em full penetration} state $J_{\parallel}=J_{c\parallel}\;,\;\forall\; z$ is reached. On the other hand, a very interesting feature is to be recalled for the paramagnetic case (lower pane of Fig. \ref{fig_14}). For the first time along the exposition of this paper we have met a set of conditions that produce an excursion of $J_{\perp}$, i.e.: $J_{\perp}=J_{c\perp}$ is violated during the process of increasing $H_{y0}$. To be specific, $J_{\perp}$ starts from the condition $J_{\perp}=J_{c\perp}$, given by the initial process in $H_{x0}$. Then, a basically linear decrease from some inner point toward the surface occurs, with an eventual reduction to a nearly null value at some regions within the sample (C-states are basically provoked). Further increase of $H_{y0}$ produces a new CT-state. This behavior is shown in a pictorial form within the insets of Fig. \ref{fig_14}. Recall that the average current density sharply transits from a T-state ($J_{\perp}=J_{c\perp}\,,\,J_{\parallel}=0$) to the CT-state ($J_{\perp}=J_{c\perp}\,,\,J_{\parallel}=J_{c\parallel}$) for the diamagnetic case, while a T $\to$ C $\to$ CT evolution happens for the initial paramagnetic conditions. This behavior allows a physical interpretation in terms of the evolution of the magnetic field profiles. Thus, as stated before, the cases with small $\chi$ are characterized by a nearly frozen profile in $H_x$, as shown in Fig. \ref{fig_13}. Then the structure of $H_{x}(z)$ and $H_{y}(z)$ is basically a cross between two straight lines. The crossing point coincides with the minimum in $J_{\perp}(z)$. Recalling the interpretation of the perpendicular component of the current density $J_{\perp}=dH/dz$, the minima should be expected as $H_{x}^{2}+H_{y}^{2}$ has a very small variation around the crossing point of the two families of nearly parallel lines.

%
%

The details about the behavior of $J_{\parallel}$ for the larger values of $\chi$ are presented in Fig. \ref{fig_15}, that corresponds to the case $\chi^{2}=7.5$. Again, owing to the complexity of the structure, we introduce the notation cT$^{(1)}$, cT$^{(2)}$ and cT$^{(3)}$, that is explained below. Let us first recall that the corner appears when the {\em partial penetration}  regime cT$^{(3)}$ extinguishes and the full sample ($0<z<d/2$) satisfies $J_{\perp}=J_{c\perp}$ and $J_{\parallel}=J_{c\parallel}$ (i.e.: CT). This property is clearly seen in the lower panel of the figure. Thus, the cT$^{(1)}$ regime is characterized by a T region in the inner part of the sample ($J_{\perp}=J_{c\perp}$ and $J_{\parallel}=0$), that abruptly becomes CT at a point that progressively penetrates toward the center (T-CT structure). At a certain instant, the profile becomes T-CT-T because the outermost layers develop a {\em subcritical} $J_{\parallel}$. This is called cT$^{(2)}$. Then, the central CT band grows toward both ends. In first instance, the inner T region becomes CT, giving a global CT-T structure, that we call cT$^{(3)}$. In a final step, the surface T layer shrinks again to a null width and the full profile is a CT region. This moment establishes the appearance of the corner in the magnetization curves.

%
%
\subsection{Infinite slab: other CS models}

As stated before, our theory will be used to investigate the properties related to several modifications of the {\em conventional} DCSM considered in the previous section. Such modifications can be justified as corrections to the simplifying ideas that flux depinning is only related to $J_{\perp}$ and that flux cutting is only related to $J_{\parallel}$. As indicated in Ref. \onlinecite{brandt_unusual}, in a general scenario, one should consider the dependencies $J_{c\perp}=J_{c\perp}(J_{\parallel})$ and $J_{c\parallel}=J_{c\parallel}(J_{\perp})$. In this work, we do not attempt a microscopic justification on how the DCSM hypotheses should be corrected. However, on the basis of minimum complexity, we will analyze two facts: (i) the flux cutting criterion will be revised so as to account for the 3--D nature of the problem. In fact, we will show that if one considers a critical angle threshold, the cutting barrier depends both on $J_{\parallel}$ and $J_{\perp}$. (ii) Also, we will investigate a smooth version of the DCSM in which the corners of the rectangular region have been rounded (see  Fig. \ref{fig_4}). Physically, the idea behind this property is that the mechanisms of flux depinning and cutting are not fully independent after all, as one could expect in a continuum theory. Along this line, we recall that an {\em elliptic} model was introduced in Ref. \onlinecite{romero} that produces a rather good description of experimental data for situations in which $J_{\perp}$ and $J_{\parallel}$ occur.

\subsubsection{Critical angle gradient in 3--D configurations}

In this part, we present some results related to the concept of
critical angle gradient in 3--D systems. It is well known that, in
fact, the limitation on $J_{\parallel}$ appears as related to the
energy reduction by the cutting of neighboring flux lines when they
are at an angle beyond some critical value.\cite{brandtjltp79,clemyeh} This
concept has been largely exploited in the 2--D slab geometry for
fields applied parallel to the surface\cite{dcsm} and is introduced by the
local relation
\begin{equation}
\left|\frac{d\alpha}{dz}\right|=\left|\frac{J_{\parallel}}{H}\right|\leq K_{c} \, ,
\end{equation}
that establishes a critical angle gradient. Here, $\alpha$ stands
for the angle between the flux lines and a given reference within
the $XY$ plane (i.e.: an azimuthal angle). However, for the 3--D cases
under consideration, the relative disorientation between flux lines
may also have a polar angle contribution, i.e.: ${\bf H}$ does not
necessarily lie within the $XY$ or any other given plane. As sketched in
Fig. \ref{fig_3}, one has to introduce the angle $\psi$ within the
plane defined by the pair of flux lines under consideration. After
some vector algebra, it can be shown that, for the infinite slab geometry, with a 3--D magnetic field one has
\begin{equation}
\label{eq:kcangle}
\frac{d\psi}{dz}=\sqrt{\frac{J_{\parallel}^{2}}{H^{2}}+\frac{H_{z}^{2}J^{2}}{H^{4}}} =\frac{1}{H}\sqrt{J_{\parallel}^{2}+\frac{H_{z}^{2}}{H^{2}}\left(J_{\parallel}^{2}+J_{\perp}^{2}\right)}\, ,
\end{equation}
where the third component is also introduced. 

The above result is just a
particular case of the relation

\begin{eqnarray}
\btd \times \left( B \hat {\bf B}\right)&&=\left[\left(\btd B\right) \times \hat {\bf B}\right] +
\left[ B\;\left(\btd\times\hat {\bf B}\right)\right]
\nonumber\\
\nonumber\\
&&\equiv \left[{\bf J}_{\perp ,1}\right]+\left[{\bf J}_{\perp ,2}+{\bf J}_{\parallel}\right]\, ,
\end{eqnarray}

showing that, in general, both ${\bf J}_{\parallel}$ and ${\bf J}_{\perp}$ can contribute to the spatial
variation of the direction $\hat{\bf B}$.

%
%
\begin{figure}
\begin{center}
\includegraphics[width=0.48\textwidth]{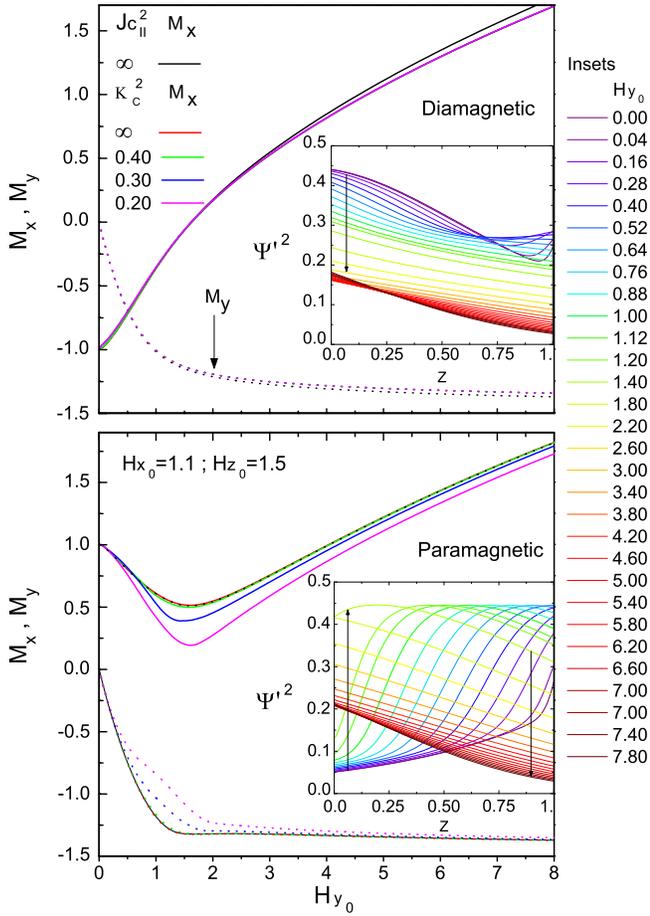}
\caption{\label{fig_16} (color online) The magnetic moments $M_{x}$
(solid lines) and $M_{y}$ (dotted lines) of the slab as a function of $H_{y_{0}}$ for the critical angle gradient model [Eq. (\ref{eq:modelKc})]. The unrestricted case ($\kappa_{c}^{2}\to\infty$) is shown for comparison with several cases with
restricted angle gradient: $\kappa_{c}^{2}=$0.20, 0.30 and 0.40 (dimensionless units are defined by $\kappa_{c}\equiv K_{c}d/2$). Shown are the
diamagnetic (top) an paramagnetic (bottom) cases for $H_{x_{0}}=1.1$
and $H_{z_{0}}$=1.5. The insets detail the evolution of the angle gradient profiles
for $\kappa_{c}^{2}\to\infty$.}
\end{center}
\end{figure}

Below, we display the effects of using the cutting limitation
\begin{equation}
\label{eq:modelKc}
\left|\frac{d\psi}{dz}\right|\leq \kappa_{c}
\end{equation}
instead of assuming a constant value for the parallel
critical current. Fig. \ref{fig_16} contains the main results. The calculations have been performed for the same magnetic processes (dia- and paramagnetic) considered in the previous section.

In general, (compare Figs. \ref{fig_9} and \ref{fig_16}) one can see that the smaller values for the cutting threshold in whatever form produce the smaller magnetic moments. However, some important differences are to be quoted. On the one side, the critical angle criterion $|{\psi}'|\leq\kappa_{c}$ produces a smooth variation, by contrast to the corner structure induced by the critical current one $J_{\parallel}\leq J_{c\parallel}$. On the other hand, the effect of changing the value of $\kappa_{c}$ is much less noticeable, especially for the diamagnetic case, in which the full range of physically meaningful values of $\kappa_{c}$ produce a negligible variation.

We call the readers' attention that the above mentioned range for $\kappa_{c}$ is established by the application of Eq. (\ref{eq:kcangle}) to the initial state of the sample. Thus, if one takes $J_{\parallel}=0,\, H_{z0}=1.5,\, H_{x0}=1.1$, the squared angle gradient takes the value ${\psi}'^2 = 0.19$ and one has to use $\kappa^2 > 0.19$ in order to be consistent with the initial critical state assumed.

\subsubsection{Smooth CS models}

%
%
\begin{figure}[!]
\begin{center}
\includegraphics[width=0.48\textwidth]{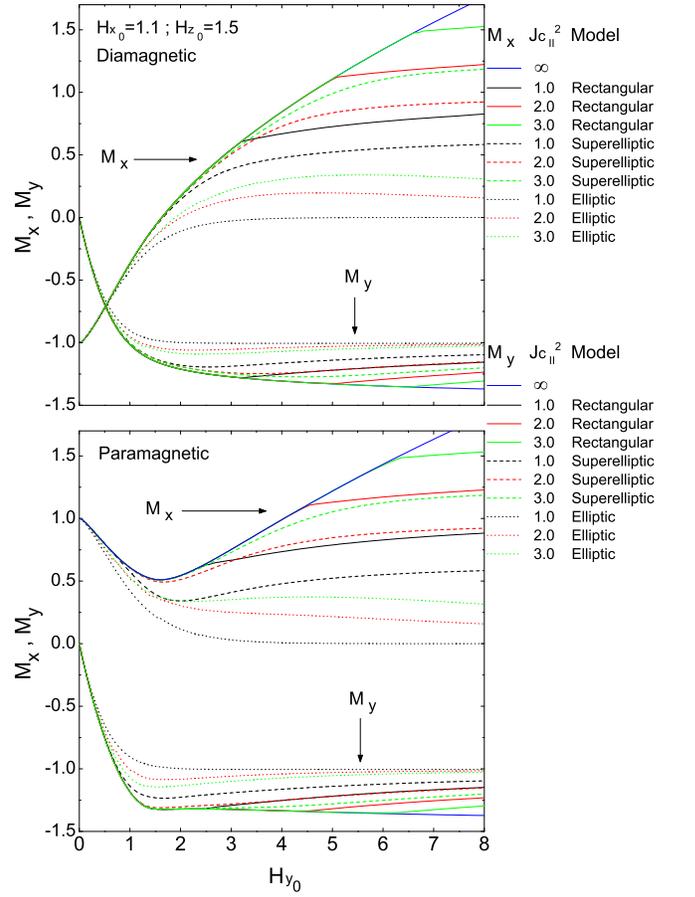}
\caption{\label{fig_17} (color online) The magnetic moments $M_{x}$
and $M_{y}$ of the slab per unit area as a function of $H_{y_{0}}$
in the diamagnetic (top) and paramagnetic (bottom) cases with
$H_{x_{0}}=1.1$ and $H_{z_{0}}$=1.5. The ``infinite band'' (external
solid lines), rectangular (solid lines), superelliptical
(dashed-lines), and elliptical (dotted-lines) models are shown for several values of the ratio $J_{c\parallel}/J_{c\perp}$.}
\end{center}
\end{figure}

Here, we develop the concept of smooth double critical state model, introduced before. Mathematically, the effect of {\em rounding the corners} for the rectangular DCSM region may be represented by a {\em one-parameter} family of functions with the generic form
\begin{equation}
\label{eq:superellipse}
\left(\frac{J_{\parallel}}{J_{c\parallel}}\right)^{2n}+
\left(\frac{J_{\perp}}{J_{c\perp}}\right)^{2n}\leq 1 \, .
\end{equation}
Such kind of curves are known as a {\em superellipses} and cover the range of interest just by allowing $n$ to take values over the positive integers. As the reader can easily verify, $n=1$ corresponds to the standard ellipse and $n\gtrsim 5$ is already a rectangle with faintly rounded corners.

\begin{figure}
\begin{center}
\includegraphics[width=0.48\textwidth]{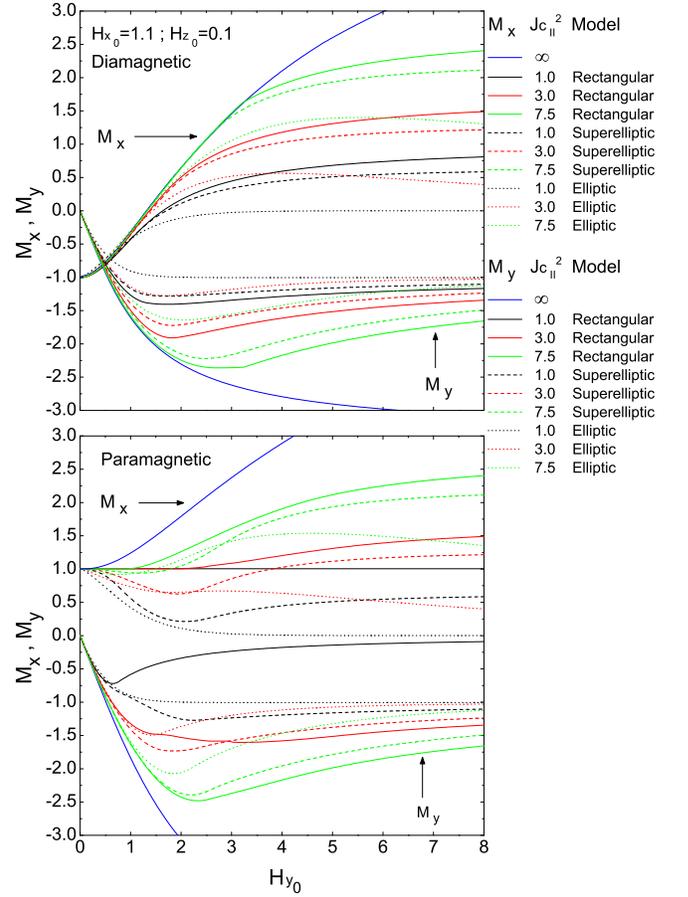}
\caption{\label{fig_18} (color online) Same as Fig. \ref{fig_17}, but now for $H_{z0}=0.1$.}
\end{center}
\end{figure}

\begin{figure}
\begin{center}
\includegraphics[width=0.48\textwidth]{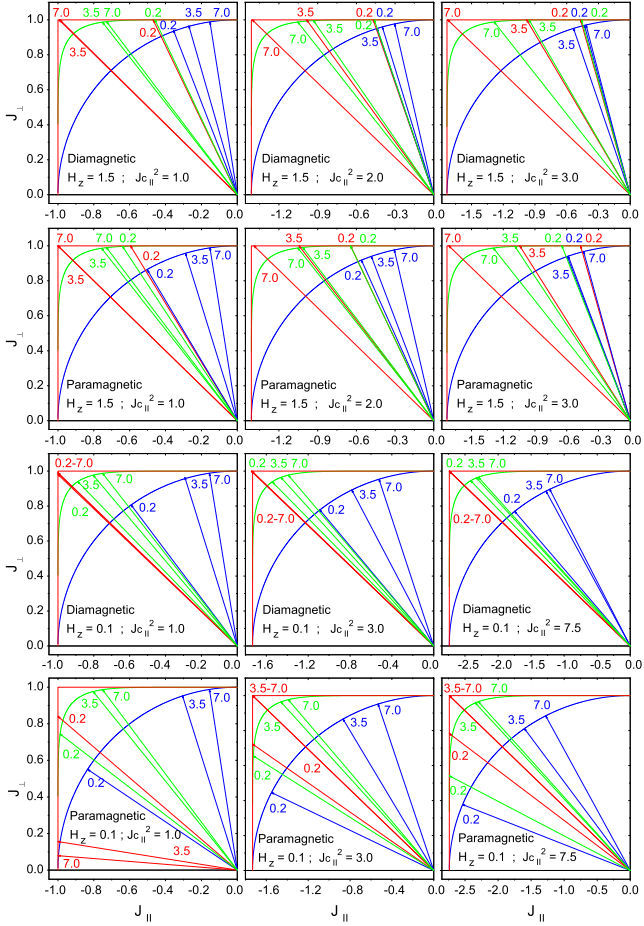}
\caption{\label{fig_19} (color online) Current density vector in the
$J_{\perp}$ vs. $J_{\parallel}$ representation for the rectangular (red),
superelliptical (green), and elliptical (blue) models. The
diamagnetic and paramagnetic cases with $H_{z0}=1.5$ and $H_{z}=0.1$ are shown for several values of the ratio $J_{c\parallel}/J_{c\perp}$ and at several values of the field $H_{y0}$ as labeled on each arrow. Recall the scales on the horizontal axes that have been re-sized for visual purposes.}
\end{center}
\end{figure}

In order to illustrate the effect of smoothing the allowed region of current density components $(J_{\parallel},J_{\perp})$, below we will show the magnetization curves that are obtained for the dia- and paramagnetic cases considered before. We compare the predictions for $n\to\infty$, $n=4$ and $n=1$. For simplicity, they will be named after rectangular, superelliptic and elliptic. The main results are plotted in Figs. \ref{fig_17}-\ref{fig_19}.

Fig. \ref{fig_17} shows the behavior of $M_x$ and $M_y$ for an external perpendicular field at the moderate intensity region $H_{z0}=1.5$. The first observation is that the overall effect of reducing the value of $\chi\equiv J_{c\parallel}/J_{c\perp}$ is the same for the three models. The smaller the value of $\chi$, the higher reduction respect to the T-state ($\chi\to\infty$) master curve for the magnetic moment components. On the other hand, as the particular details for each model, we recall: (i) the smooth models lead to smooth variations, i.e.: the corner is not present, (ii) the breakdown of the T-state behavior occurs before (at higher values of $\chi$ or lower values of $H_{y0}$) for the smoother models. Strictly speaking, the concept of T-state is only valid for the rectangular region, but it is asymptotically generated as the superellipse parameter $n$ grows. Finally, (iii) The isotropic CS limit, given by the circular region $n=1\;\&\; \chi =1$ produces the expected results:\cite{badiaPRL} $M_x$ collapses to zero, and $M_y$ develops a {\em one dimensional} critical state behavior.

%
%

Fig. \ref{fig_18} shows the comparison of $M_x$ and $M_y$ for the same models considered above, but now for a low perpendicular field ($H_{z0}=0.1$). Here, one can notice: (i) the rectangular and superelliptical models produce very similar results for the diamagnetic case, both for $M_x$ and for $M_y$, noticeably differing from the elliptical region predictions, that still show a practical collapse of $M_x$ and a saturation in $M_y$ as stated before. (ii) The paramagnetic case involves a higher complexity. Thus, we recall that the already mentioned ``flat'' behavior of $M_{x}$ for small values of $\chi$ within the rectangular region model. This feature is no longer observed upon smoothing of the restriction region. On the contrary, the smooth models involve an initial negative slope and a minimum, resembling the behavior of $M_x$ for the rectangular model, but in moderate $H_{z0}$. As concerns $M_y$, important differences among the three models are also to be recalled.

%
%

In order to provide a physical interpretation of the behaviors reported in the above paragraphs, a comparative plot of the current density vectors for each case is given in Fig. \ref{fig_19}. For clarity, we restrict to the representation of the vector ${\bf J}$ at the surface of the sample ($z=d/2$) for a selected number of values of $H_{y0}$. Just at a first glance, one can relate the  best coincidence in predicted magnetization to the more similar critical current density structures (superelliptical and rectangular regions for the diamagnetic case with $H_{z0}=0.1$). Recall that, in this case, the rectangular region produces a CT-state structure
($J_{\parallel}=J_{c\parallel}$ and $J_{\perp}=J_{c\perp}$) that is represented by a ${\bf J}$ vector, pinned in the corner. On the other hand, the vector {\bf J} related to the superelliptic model does not pin at any point, because such a singular point does not exist. However, it is basically oriented in the same fashion and this relates to the good agreement in ${\bf M}$. We finally emphasize that the cases in which strong differences occur for the magnetic moment, are also related to important changes in the behavior of ${\bf J}$. Thus, if one considers the paramagnetic case at small values of $H_{z0}$ and $H_{y0}$, the significant differences in magnetization relate to an opposite behavior in ${\bf J}$. Moreover (see left bottom panel of Fig. \ref{fig_19}), the rectangular model predicts a transition toward a C-state ($J_{\parallel}=J_{c\parallel}$ and $J_{\perp}\approx 0$), while the smooth versions produce a tendency toward the T-state.

%
%
\subsection{The longitudinal problem with transport currents.}
\label{longtransp}

In this section, we still investigate the infinite slab geometry, now
under the assumption of a longitudinal transport current. Such configuration has been a longstanding problem, related to the design of superconducting devices, and is still frequently focused,\cite{asulay} and described in terms of the critical state regime. Here, we will consider the slab geometry, subjected to a uniform field normal to
the surface ($H_{z0}$), then a transport current applied along the
$y-axis$, and eventually, a magnetic field ($H_{y0}$) along the same direction. This situation matches the second
example in Ref. \onlinecite{brandt_unusual}, but here, no
restrictions will be required for the ratios $\chi^{-1}\equiv
J_{c\perp}/J_{c\parallel}$ and $\varsigma\equiv J_{c\perp}d/H_{z0}$, that
are small parameters in that case. Notice that the smallness of
$\chi^{-1}$ means that the arising critical state is approximated by the
unbounded band region $|J_{\perp}|=J_{c\perp}, 0 < |J_{\parallel}|
<\infty$ described before (T-states). The smallness of $\varsigma$ was meant to indicate a small deviation of the full magnetic field respect to the
$z-axis$. Then, moderate values of $J_{\parallel}$ are expected. Remarkably, these hypotheses allowed to obtain a set of approximate analytic formulas for the electromagnetic quantities that allow to bypass the numerical solution of the differential equations. However, as it will be shown below, the range of application is narrower than expected. By using our numerical method, that allows to calculate the sample's response for any value of the parameters $\chi$ and $\varsigma$, the range of application of such approximation will be discussed.

In brief, our results are not limited to the {\em weak longitudinal
current} conditions. By contrast, the calculations are performed
numerically, allowing to display the corrections needed in the
general critical states.

\subsubsection{Mathematical statement}

Technically, the application of a transport current relates to the
consideration of specific boundary conditions for the
electromagnetic fields. Within our mutual inductance formulation
[Eqs. (\ref{eq:fields}-\ref{eq:mcoef})], the above described longitudinal problem
takes the following form within the DCSM framework. One has to minimize

\begin{eqnarray}
\label{eq:minprintra}
{\tt F}&=&\displaystyle{\displaystyle \frac{1}{2}}\sum_{i,j}\xi_{i,{\rm n+1}}M_{ij}^{x}\,\xi_{j,{\rm n+1}}
-\sum_{i,j}\xi_{i,{\rm n}}M_{ij}^{x}\,\xi_{j,{\rm n+1}}
\nonumber\\
&+&\displaystyle{\displaystyle \frac{1}{2}}\sum_{i,j}\psi_{i,{\rm n+1}}M_{ij}^{y}\,\psi_{j,{\rm n+1}}
-\sum_{i,j}\psi_{i,{\rm n}}M_{ij}^{y}\,\psi_{j,{\rm n+1}}
\nonumber\\
&+&\sum_{i}\xi_{i,{\rm n+1}}(i-1/2)(H_{y0,{\rm n+1}}-H_{y0,{\rm n}})
\end{eqnarray}

for

\begin{eqnarray}
\label{eq:jpapetra}
(1-h_{x,i}^{2})\xi_{i}^{2}+(1-h_{y,i}^{2})\psi_{i}^{2}-2h_{x,i}h_{y,i}\,\xi_{i}\psi_{i}\leq j_{c\perp}^{2}
\nonumber\\
\nonumber\\
h_{x,i}^{2}\,\xi_{i}^{2}+h_{y,i}^{2}\,\psi_{i}^{2}+2h_{x,i}h_{y,i}\,\xi_{i}\psi_{i}\leq j_{c\parallel}^{2} \, 
\end{eqnarray}

and

\begin{equation}
\label{eq:transportcurrent}
\sum_{i}\psi_{i}=I_{\rm transport} \, .
\end{equation}

This last condition, indicates that a certain transport current is
being applied to the sample.

\begin{figure}[t]
\includegraphics[width=0.5\textwidth]{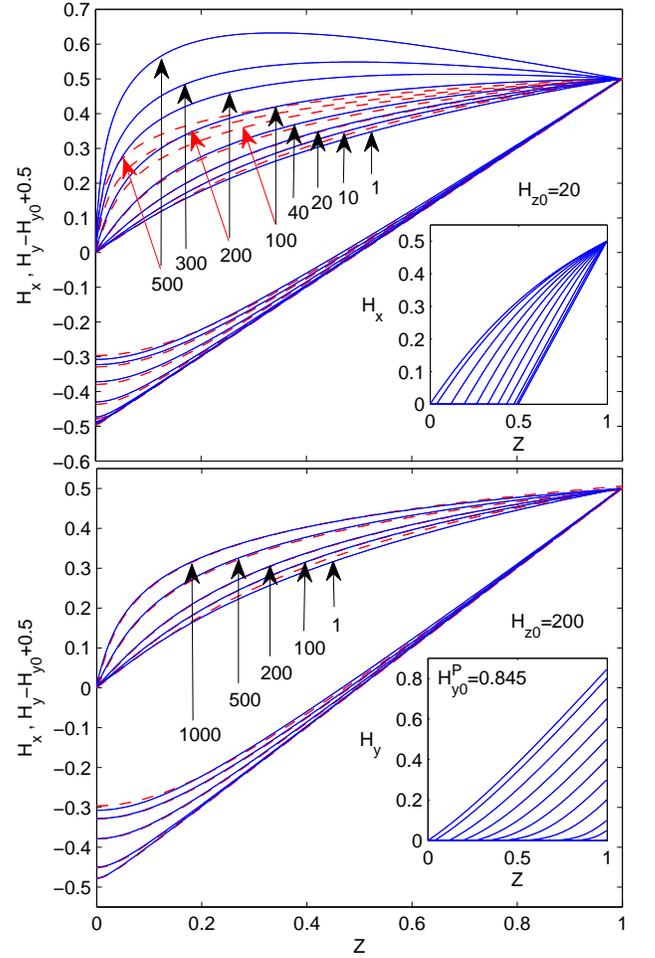}
\caption{\label{fig_20} (color online) Profiles of the magnetic field components $H_x(z)$ and $H_y(z)$ for the longitudinal problem corresponding to a transport current along the $y$ axis of value $I_{\rm transport}=J_{c\perp}d/2$ and at several increasing values of the magnetic field $H_{y0}$ as labeled in the curves. A slab geometry with uniform perpendicular field ($H_{z0}=20$ and then $H_{z0}=200$) was assumed. The plot shows the comparison of the full range numerical solution (continuous lines) to the analytical approximation in Eq. (\ref{eq_Brandt_aprox}) (dashed). The insets show the initial flux penetration profiles for both components of the magnetic field. The infinite band (T-state) model has been assumed.}
\end{figure}

\begin{figure*}[!]
\begin{center}
\includegraphics[width=1\textwidth]{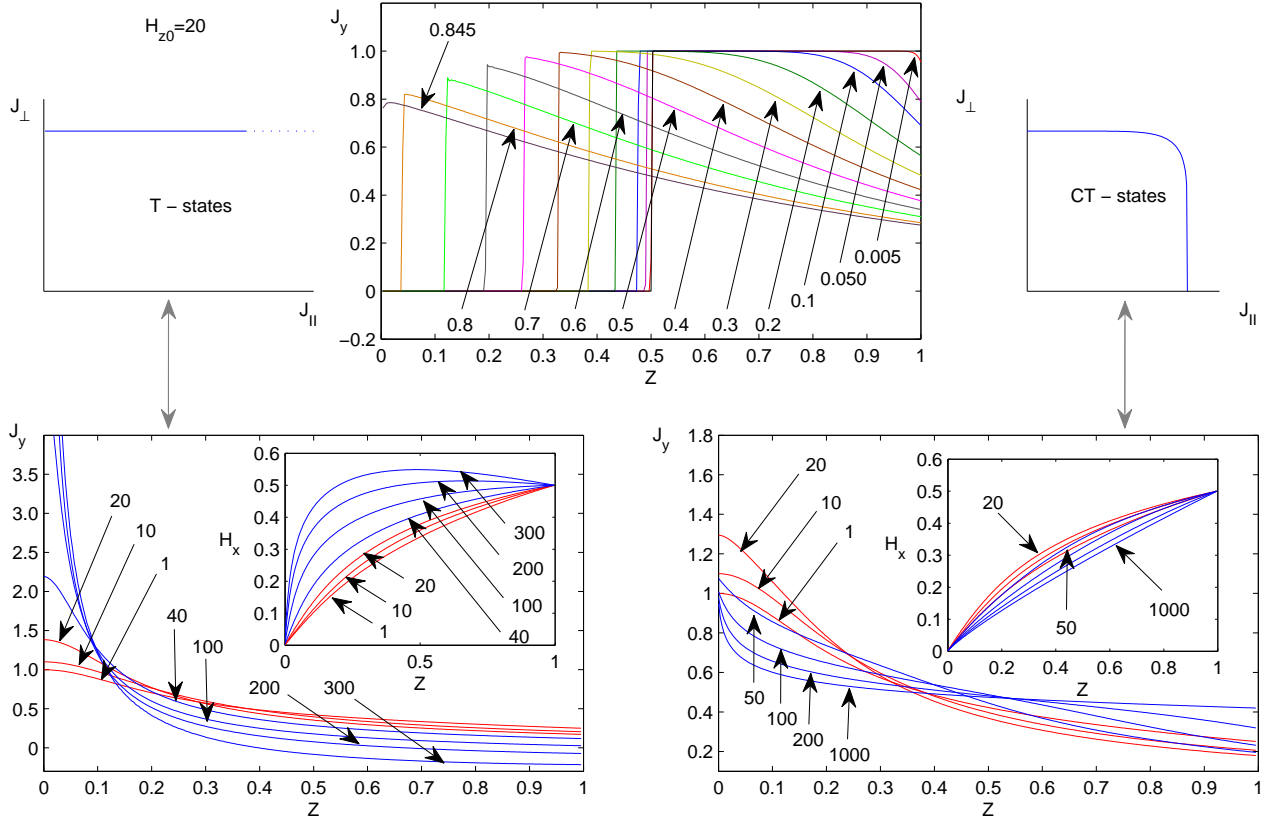}
\caption{\label{fig_21} (color online) Profiles of the transport current density $J_{y}(z)$ for the longitudinal transport problem in the slab for the same conditions of Fig. \ref{fig_20} (here $H_{z0}=20$). The upper panel shows the initial partial penetration process induced by application of increasing values of $H_{y0}$ as labeled in the successive curves. The transport current profile fully penetrates at $H_{y0}^{p}=0.845$. The lower panels show the evolution of $J_{y}(z)$ from the fully penetrated state $H_{y0}>1.0$ both for the T-state model (left) and for a superelliptical model with $n=4$ and $\chi =1$ (right). The induced magnetic field profiles $H_x$ are shown as insets.}
\end{center}
\end{figure*}

On the other hand, as related to the symmetry properties for the
transport configuration [$\psi_{i}(z)=\psi_{i}(-z)$ as opposed to
the antisymmetry for the case of shielding currents], here one has
to use the mutual inductance expressions

\begin{eqnarray}
\label{eq:mcoeftransp}
M_{ij}^{x}\equiv 1+2\left[{\rm min}\left\{ i,j\right\}\right]
\nonumber\\
M_{ij}^{y}\equiv 1+2\left[N-{\rm max}\left\{ i,j\right\}\right]
\nonumber\\
M_{ii}^{x}\equiv 2\left(\frac{1}{4}+i-1\right)
\nonumber\\
M_{ii}^{y}\equiv 2\left(\frac{1}{4}+N-i\right)
\end{eqnarray}
with $N$ the full number of layers in the discretized slab. As a
final detail, the evaluation of the magnetic fields has to be made
according to
\begin{eqnarray}
\label{eq:fieldstra}
&&H_{x,i}=\sum_{j<i}\psi_{j}+\psi_{i}/2
\nonumber\\
&&H_{y,i}=\sum_{j>i}\xi_{j}+\xi_{i}/2 \, .
\end{eqnarray}
The results obtained by application of
Eqs. (\ref{eq:minprintra}--~\ref{eq:fieldstra}) are displayed in
Figs. \ref{fig_20} and \ref{fig_21}. They are described below.

%
%

\subsubsection{T-state solutions}

%
%

First, we analyze the case in which the critical current ratio
$J_{c\perp}/J_{c\parallel}$ is small, i.e.: T-states are warranted.
Nevertheless, here, the existence of moderate or even high values for the
parallel component of $\bf J$ will be allowed.

Fig. \ref{fig_20} shows the main features of our investigation. This
includes the comparison of the penetration profiles for $H_x$ and
$H_y$ obtained from our theory and from the analytic expressions in
Ref. \onlinecite{brandt_unusual}, i.e.:
\begin{eqnarray} \label{eq_Brandt_aprox}
H_{x}=&&\frac{\alpha}{{\rm cos}\,\theta}{\rm arcsinh}\left(\frac{z}{\alpha}\right)
\nonumber\\
H_{y}=&&H_{y0}-\alpha\left(\sqrt{1+\frac{1}{\alpha^{2}}}-\sqrt{1+\frac{z^2}{\alpha^2}}\right)
\nonumber\\
{\rm cos}\,\theta =&& 2\alpha\,{\rm arcsinh}\left(\frac{1}{\alpha}\right) \, .
\end{eqnarray}
Here, $\alpha$ has to be obtained for each value of the applied field from the condition $cos\theta=H_{z0}/\sqrt{H_{z0}^{2}+H_{y0}^{2}}$.

One can notice that the agreement
is rather good  for the higher value of $H_{z0}$ (200 in our
dimensionless units), whereas remarkable differences appear for
$H_{z0}=20$ as $H_{y0}$ increases. Our
interpretation of the facts is as follows. 

As regards the establishment of the full penetration profile, we have
straightforwardly obtained this condition through the step-by-step
integration starting from the state $H_{y0}=0$ (the evolution is shown in the
insets of the figure). Whereas the value 0.796 is estimated for the
penetration field $H_{y0}^{p}$ within the analytical limit, by the straightforward method described above we get
$H_{y0}^{p}=0.845$. In spite of some small differences for the low field profiles, at moderate values ($H_{y0}\lesssim H_{z0}$) the curves always coincide. On the other
hand, the failure of the analytical approximation for the higher
values of $H_{y0}$ is readily explained by the observation of the
plot. Thus, increasing $H_{y0}$ can compress the transport current
toward the center of the sample (as indicated by the slope of
$H_{x}(z)$). For the case of $H_{z0}=20$, one gets $J_{y,max}\approx 5$
when $H_{y}\approx 100$ and $J_{y,max}\approx 50$
when $H_{y}\approx 1000$, then a considerable value of $J_{\parallel}$ is obtained. This leads to a not so good approximation from the analytic condition in the approximation of Ref. \onlinecite{brandt_unusual}, which one is only valid for small values of this quantity. However, when comparison is made
for $H_{z0}=200$, one gets $J_{y,max}\approx 1$
when $H_{y}\approx 100$ and $J_{y,max}\approx 5$
when $H_{y}\approx 1000$. Then, a much better performance is obtained for the analytical limit even for very high applied fields $H_{y0}$.

\subsubsection{CT-state solutions}

This part will be devoted to unveil the features of longitudinal
transport problems under general critical state conditions. To be
specific, we will compare the transport current profiles obtained
either by assuming a T-state (flux cutting may be neglected) or a
general critical state in which flux pinning and transport are at
the same level, i.e.: $\chi=J_{c\parallel}/J_{c\perp}=1$. Pictorially, (see
Fig. \ref{fig_21}), we solve the CS problem either for a horizontal
band or for a smoothed square region. All the results shown
correspond to the case $H_{z0}=20$.

First, let us recall that the initial transport profile ($J_{y}(z)=0$
for $z < 0.5$ and $J_{y}(z)=1$ for $z\ge 0.5$) is spread out by the
action of $H_{y0}$. Thus, as illustrated in the upper panel of
Fig. \ref{fig_21}, the free current core is reduced step-by-step
until the eventual full penetration occurs for an applied field
$H_{y0}^{p}=0.845$. We remark that no significant difference is observed
when this plot is generated, either for the T or CT critical
state models. On the other hand, the full penetration regime
displays clear differences, at least for the high field region.
Thus, for the case under consideration ($H_{z0}=20$) one can notice
that the transport current profiles are very similar for $H_{y0}\leq
20$, but display even qualitative differences for $H_{y0}> 20$. This
property relates to the appearance of the limitation for
$J_{\parallel}$. In fact, the value $J_{\parallel}=0.9
J_{c\parallel}=0.9 J_{c\perp}$ is obtained at $z=0$ when $H_{y0}$ equals 20.
Subsequently, $J_{\parallel}$ (which one can basically identify with
the transport current $J_y$ for large values of $H_y$) increases
more and more at the central region of the sample for the T-state model. For the
CT-state, the situation at high fields is rather
different. Initially, $J_y$ reaches the limit $J_{c\parallel}$ ($=J_{c\perp}$ in the case under study) at the center,
and rapidly decreases toward the value $J_{c\perp}/2$, that is roughly
maintained for all $z>0$. 

The behavior of ${\bf J}$ is obviously inherited by the flux profiles. One can see it 
in the plot of $H_x$ (see the insets of Fig. \ref{fig_21}). For the
T-states, $H_x$ saturates at the periphery, where a practically flat profile is reached. Then, one has $J_y\lesssim 0$ in that region, and the full transport profile is {\em shifted} toward the center of the sample. At the same time, the shielding part of the critical current $J_x\approx J_{\perp}$ is maximum within the region of negligible transport and goes to zero where transport predominates. Such a behavior (compression of transport by penetrating shielding currents) is straightforwardly deduced from the slopes of the magnetic field profiles in Fig. \ref{fig_20} and has been already suggested as a possible explanation the magnetic field dependence of the transport critical current in longitudinal geometry.\cite{baltaga} Direct measurements of the transport current density profile,\cite{voloshin} have also been used to conclude that a longitudinal field compresses the transport current toward the center of the sample. On the other side, for the CT-state case, in which the limitation on the parallel current density is active, 
$H_x$ displays a turn back until a nearly linear penetration is
reached, basically characterized by the slope $J_{c\perp}/2$. Thus, the transport profile eventually {\em stretches} instead of concentrating toward the center.

%
%
%
\section{Conclusions}
\label{conclusions}

In this article, we have shown that the {\em critical state theory} for the magnetic response of type-II superconductors, may be built in a very general framework. The basic concepts underlying the phenomenological approach, issued by C. P. Bean in the early sixties have been identified as

\begin{enumerate}
\item The CS theory bears a Magneto Quasi Stationary approximation for the Maxwell equations in which $\dot{\bf E}$ and $\dot{\rho}$ are second order quantities. This means that the {\em magnetic flux dynamics} is described by

\begin{eqnarray}
\left\lgroup\qquad
\begin{array}{ll}
\nabla\times{\bf B} &= \mu_{0}{\bf J}
\nonumber\\
\Inc{\bf B} &= \nabla\times ({\bf E}\Inc t)\,\qquad {\rm (implicit)}
\nonumber
\end{array}
\qquad\right\rgroup
\end{eqnarray}

i.e.: the inductive part of ${\bf E}$ may be introduced through Faraday's law, whereas the role of electrostatic quantities is irrelevant. {\bf E} may be modified by a gradient (${\bf E}\to{\bf E}+\nabla\phi$) with no effect on the magnetic response.

\item The law that characterizes the {\em conducting behavior} of the material may be written in the form

\begin{eqnarray}
\left\lgroup\qquad
\begin{array}{ll}
{\rm If}\quad E = 0 \quad &\Rightarrow \quad \dot{\bf J} = 0
\nonumber\\
{\rm If}\quad E \neq 0 \quad &\Rightarrow \quad {\rm max}\,{\bf J}\cdot\hat{\bf E}\left.\right|_{{\bf J}\in\region}
\nonumber
\end{array}
\qquad\right\rgroup
\end{eqnarray}
\end{enumerate}

In physical terms, the material ``reacts'' with a maximal shielding rule when electric fields are induced. A perfect conducting behavior characterizes the magnetostatic equilibrium when external variations cease. In all cases, ${\bf J}$ is constrained within some region $\region$.

We stress that the importance of ${\bf E}$ is sometimes veiled by the actual application of the above rules, as it plays an implicit role usually.

On the other hand, the above representation may be understood as the macroscopic counterpart of the underlying vortex physics. Thus, the physical barriers for flux depinning and cutting are represented by the condition ${\bf J}\in\region$, i.e.: the current density is confined within some region $\region$ ($J\leq J_{c}$ in 1--D). The evolution from one magnetostatic configuration to another occurs through the local violation of this condition, i.e.: ${\bf J}\notin\region$ ($J>J_{c}$ in 1--D problems). However, owing to the high dissipation, an almost instantaneous response may be assumed, represented by a {\em maximum shielding} rule in the form ${\rm max}\,{\bf J}\cdot\hat{\bf E}\left.\right|_{{\bf J}\in\region}$ ($J=\pm J_{c}$ in 1--D).   

The general CS theory exposed above may be solved in different forms. In our work, we emphasize the performance of variational methods for solving the problem. In particular, the mutual inductance representation with ${\bf J}({\bf r})$ as the unknown, offers two important advantages: (i) intricate boundary conditions and infinite domains are avoided, (ii) the transparency of the numerical statement and its performance (stability) are outlined. Thus, the quantities of interest (flux penetration profiles and magnetic moment) are obtained by integration and additional smoothing is ensured. To be specific, upon discretization, the CS problem bears the algebraic expression

\begin{eqnarray}
\left\lgroup\;
\begin{array}{ll}
{\tt min\quad F}&={
{\displaystyle\frac{1}{2}}}\sum_{i,j}I_{i,{\rm n+1}}M_{ij}\, I_{j,{\rm n+1}}
\nonumber\\
\nonumber\\
&-\sum_{i,j}I_{i,{\rm n}}M_{ij}\, I_{j,{\rm n+1}}
+\sum_{i}I_{i,{\rm n+1}}\Inc H_{i}
\nonumber
\end{array}
\;\right\rgroup
\end{eqnarray}

with $\{I_{i,{\rm n+1}}\}$ the set of unknown current values at the specific circuits for the problem of interest, $M_{ij}$ their {\em mutual inductance} coupling coefficients, and $\Inc H_{i}$ the applied magnetic field increment. Corresponding to the CS rule ${\bf J}\in\region$, each value $I_i$ must be constrained. Also, we have found that a number of constraints related to physically meaningful CS models may be expressed in the algebraic form  

\begin{equation}
\left\lgroup\qquad
{\textstyle{F}}_{\alpha}\bigl({\Sigma}_{j}\; I_{i}C_{ij}^{\alpha}I_{j}\bigr)\leq f_{0\alpha} \quad\forall i
\nonumber
\qquad\right\rgroup
\end{equation}

with $f_{0}$ some constant representing the physical threshold, $F_{\alpha}$ an algebraic function representing the physical model, and $C_{ij}^{\alpha}$ a coupling matrix, that also depends on the model. 
For example, the isotropic model corresponds to $F(x)=x\; ;\; C_{ij}=\delta_{ij}\; ; \; f_{0}={\jmath}_{c}^{2}$. On the other hand, the double critical state model is given by
\begin{eqnarray}
F_{1}(x)=x\quad ; \quad F_{2}(x)=x
\nonumber\\
f_{01}={\jmath}_{c\perp}^{2}\quad ; \quad f_{02}={\jmath}_{c\parallel}^{2}
\nonumber
\end{eqnarray}
and the coupling coefficients $C_{ij}^{1,2}$ that project the local current density onto the local magnetic field or its normal plane, are obtained from Eqs. (\ref{eq:fields}) and (\ref{eq:jpape}).

General critical state problems have been solved for a number of examples, within the infinite slab geometry. All of them share a 3--D configuration for the magnetic field, i.e.: ${\bf H}=(H_{x},H_{y},H_{z})$ under various magnetic processes and models for the critical current restriction. Thus, we have considered several physical scenarios classified by the ansatz for the flux depinning and cutting processes (basically affecting the critical current thresholds $J_{c\perp}$ and $J_{c\parallel}$) and their relative importance (given by $\chi\equiv J_{c\parallel}/J_{c\perp}$). In summary, the following cases have been analyzed:

\begin{enumerate}

\item T-state solutions, in which the approximation $~\chi~\gg~1$ produces the result $J_{\perp}=J_{c\perp}$ and $J_{\parallel}$ may  be arbitrarily high. Our predictions show an excellent agreement with previous results in the literature, and extend the theory to the full range of applied magnetic fields.

\item CT-state solutions in which $\chi\gtrsim 1$ for several cases within the {\em rectangular region} given by $J_{\perp}\leq J_{c\perp}$ and $J_{\parallel}\leq J_{c\parallel}$ are predicted by the theory. Outstandingly, the appearance of the flux cutting limitation takes place as a sudden corner in the magnetic moment curves in many cases. The corner establishes a criterion for the range of application of T-state models.

\item The critical angle (between vortices) criterion that establishes the limitation on $J_{\parallel}$ has been modified for 3--D problems. It is shown that, in general, the concept may involve both $J_{\parallel}$ and $J_{\perp}$ as one can see in Eqs. (\ref{eq:kcangle},\ref{eq:modelKc}).

\item The possible coupling between the flux depinning and cutting limitations has been studied through the solution of {\em smoothed} DCSM cases. In particular, we consider the effect of rounding the corners of the rectangular region $J_{\perp}\leq J_{c\perp}$ and $J_{\parallel}\leq J_{c\parallel}$, by the {\em superelliptic} region criterion $(J_{\parallel}/ J_{c\parallel})^{2n}+(J_{\perp}/ J_{c\perp})^{2n}\leq 1$ with $1\leq n < \infty$. It is shown that, under specific conditions (paramagnetic initial state and low perpendicular fields), important differences in the predictions of the magnetic moment behavior are to be expected. The differences in ${\bf M}$ have been related to the behavior of the critical current vector ${\bf J}_c$ around the corner of the rectangular region.

\item The longitudinal transport problem, i.e.: a magnetic field is applied parallel to the transport current, has been studied for several 3--D configurations. It is shown that the transport current is essentially {\em compressed} toward the center of the sample by the effect of shielding currents when no limitation on $J_{\parallel}$ is active (T-states). However, increasing the parallel field when the constraint $J_{c\parallel}$ is reached, produces a flattening on the transport current density that becomes nearly uniform across the sample.

\end{enumerate}

We emphasize that the scope of our theory is rather beyond the actual examples treated in this article. On the one side, we have shown that the CS concept allows arbitrariness in the presence of electrostatic charge and potential, and one could simply upgrade the models by the rule ${\bf E}\to{\bf E}+\nabla\phi$ if necessary. For instance, a scalar function  $\phi$ may be introduced if the direction of ${\bf E}$ has to be modified respect to the maximum shielding rule in the MQS limit.

On the other side, the extension of the theory to arbitrary sample geometries is intrinsically allowed by the mutual inductance representation. This article has laid necessary groundwork for attacking general critical state problems in 3--D geometry. Experimental studies that could reproduce the situations considered in the different examples, as a means of testing the {\em double critical state model} predictions are suggested.

From the theoretical point of view, a relevant technical issue to be considered is that the divergenceless character of the current density is not always ensured. In this work, the problem's symmetry has allowed to identify the elementary current circuits that fulfill such condition (infinite horizontal layers), and the corresponding geometrical problem of finding their coupling matrix coefficients has been solved. In general, this is not a trivial issue and one will have to incorporate the additional restriction $\nabla\cdot{\bf J}=0$ or use a representation with appropriate basis functions for ${\bf J}$.\cite{albanese}

\section*{Acknowledgment}

This work was supported
by Spanish CICyT projects MAT2008-05983-C03-01 and MTM2006-10531. H. S. Ruiz acknowledges a grant from the Spanish CSIC (JAE program).

%
%

\end{document}